\documentclass[]{aa}  

\usepackage{graphicx}
\usepackage[dvipsnames]{xcolor}
\usepackage{xfrac}
\usepackage{txfonts}
\begin{document}

    \title{Strong size evolution of disc galaxies since z=1}

    \subtitle{Readdressing galaxy growth using a physically motivated size indicator \thanks{Table \ref{tab:sample_params} are only available in electronic form at the CDS via anonymous ftp to cdsarc.u-strasbg.fr (130.79.128.5) or via http://cdsweb.u-strasbg.fr/cgi-bin/qcat?J/A+A/.}}

    \author{Fernando Buitrago
           \inst{1,2} 
           \and
           Ignacio Trujillo
           \inst{3,4}
           }

    \institute{
    Departamento de F\'{i}sica Te\'{o}rica, At\'{o}mica y \'{O}ptica, Universidad de Valladolid, 47011 Valladolid, Spain\\
    \email{fbuitrago@uva.es}
    \and
    Instituto de Astrof\'{i}sica e Ci\^{e}ncias do Espa\c{c}o, Universidade de Lisboa, OAL, Tapada da Ajuda, PT1349-018 Lisbon, Portugal
    \and
    Instituto de Astrof\'{i}sica de Canarias, V\'{i}a L\'{a}ctea $s\backslash n$, 38200 La Laguna, Tenerife, Spain
    \and
    Departamento de Astrof\'{i}sica, Universidad de La Laguna, E-38205, La Laguna, Tenerife, Spain
    }

    \date{Received September 15, 1996; accepted March 16, 1997}


\abstract{Our understanding of how the size of galaxies has evolved over cosmic time is based on the use of the half-light (effective) radius as a size indicator. Although the half-light radius has many advantages for structurally parameterising galaxies, it does not provide a measure of the global extent of the objects, but only an indication of the size of the region containing the innermost 50\% of the galaxy's light.  Therefore, the observed mild evolution of the effective radius of disc galaxies with cosmic time is conditioned by the evolution of the central part of the galaxies rather than by the evolutionary properties of the whole structure. Expanding on recent works, we studied the size evolution of disc galaxies using the radial location of the gas density threshold for star formation as a size indicator. As a proxy to evaluate this quantity, we used the radial position of the truncation (edge) in the stellar surface mass density profiles of galaxies. To conduct this task, we selected 1048 disc galaxies with M$_{\rm stellar}$ $>$ 10$^{10}$ M$_{\odot}$ and spectroscopic redshifts up to z=1 within the HST CANDELS fields. We derived their surface brightness, colour and stellar mass density profiles. Using the new size indicator, the observed scatter of the size-mass relation ($\sim$0.1 dex) decreases by a factor of $\sim$2 compared to that using the effective radius. At a fixed stellar mass, Milky Way-like (MW-like; M$_{\rm stellar}$ $\sim$ 5$\times$10$^{10}$ M$_{\odot}$) disc galaxies have, on average, increased their sizes by a factor of two in the last 8 Gyr, while the surface stellar mass density at the edge position ($\Sigma_{\rm edge}$) has decreased by more than an order of magnitude from $\sim$13 M$_{\odot}$/pc$^2$ (z=1) to $\sim$1 M$_{\odot}$/pc$^2$ (z=0). These results reflect a dramatic evolution of the outer part of MW-like disc galaxies, with an average radial growth rate of its discs of about 1.5 kpc Gyr$^{-1}$.
  }

\keywords{Galaxies: evolution -- Galaxies: structure -- Galaxies: fundamental parameters -- Methods: observational -- Galaxies: star formation -- Galaxies: spiral}

\titlerunning{Strong size evolution of disc galaxies}

\authorrunning{Buitrago \& Trujillo}

\maketitle
%

\section{Introduction}

Many alternatives for measuring the size of galaxies have been proposed over the years \citep[see a recent review in][]{Chamba20_hist}. Constrained by the technological limitations of their time, many works delimited galaxy sizes by using isophotal contours, for example 25 mag/arcsec$^2$ in the B band \citep[][]{Redman1936} or 26.5 mag/arcsec$^2$ \citep[][]{Holmberg1958}. Other choices for estimating the size were based on the radius containing a certain amount of light from the object. The most favoured proposal by far has been that of \citet{deVaucouleurs1948} to use the effective radius (r$_e$, the radius containing half the light) as a size indicator for spheroidal galaxies. With the advent of wide-area surveys such as the Sloan Digital Sky Survey \citep[SDSS;][]{York2000}, the need to have an automatic and robust way to characterise the sizes of millions of galaxies resulted in the effective radius becoming the most popular (and almost exclusive) galaxy size proxy since then.

Although using the effective radius has some distinct advantages, such as its robustness against image depth (as the galaxy's light profiles decrease exponentially or even more steeply), its identification with the galaxy size can lead to confusion \citep[see e.g.][]{Papaderos1996,Breda19,Suess19,Chamba20,dosReis20}. The main problem with the effective radius is that it does not provide a measure of the global extent of the galaxy, but only characterises the size at which 50\% of the innermost light is distributed. This can therefore lead to situations where galaxies with very different effective radii (depending on whether or not there is a bright source, e.g. a bulge, bar, ring, etc., at their centre) can have the same global extent (as characterised by the position of their disc edge, for example). To overcome the problems of the previous size characterisations, \citet{Trujillo20} proposed a physically motivated definition of the size of a galaxy. According to the new definition, the size of a galaxy would be given by the farthest radial location where the gas has been efficiently able to collapse and transform into stars. In this way, this new size proposal could potentially separate the regions of the galaxy that are mostly made up of the stellar component formed in situ from the region at the outskirts, which is likely to be dominated by the accreted material \citep[see e.g.][]{Font20}. \citet{Trujillo20} show that using the new size definition, the observed dispersion in the size-mass relation decreases by a factor of about three with respect to that resulting from using the effective radius \citep[for a theoretical discussion about the meaning of a decrease in the dispersion, see e.g. ][]{Sanchez-Almeida20}.

An expected outcome of the gas density threshold for star formation is a sharp drop in the surface brightness profiles of the galaxies in the outer parts. These sudden declines on the surface brightness profiles were found in the past in edge-on disc galaxies and were dubbed as truncations \citep{vanderKruit1979,vanderKruit1981a,vanderKruit1981b}. Motivated by both theoretical expectations and observational findings which agree on a value of  $\sim$1 M$_{\odot}$ pc$^{-2}$ for the stellar mass density at the truncation position in present-day MW-like galaxies galaxies \citep{Martinez-Lombilla19, Diaz-Garcia22}, \citet{Trujillo20} proposed using the radial location where the stellar mass density profile reaches such a value as a proxy for measuring the size of the galaxies. There is no guarantee, however, that such a proxy based on MW-like galaxies should be able to characterise the  radius to which in situ star formation takes place for all galaxies regardless of their mass, morphology, environment, or evolutionary epoch. For this reason, \citet{Chamba22} explored the edges of galaxies in the local Universe for a sample of about a thousand objects covering a large range in mass and morphologies. \citet{Chamba22} found that the stellar mass density at the truncation  in these objects is moderately mass-dependent, being smaller for dwarf galaxies ($\sim$0.6 M$_{\odot}$ pc$^{-2}$) than for objects of Milky Way mass ($\sim$1 M$_{\odot}$ pc$^{-2}$). We have followed the same methodology here as in \citet{Chamba22} to identify the size of galaxies but this time with the aim of exploring their evolution with cosmic time.

The purpose of this work is to revisit the size evolution of the disc galaxies with stellar masses similar to the MW using the new physically motivated definition by \citet{Trujillo20}. Since most previous works on galaxy size evolution have used the effective radius as a size indicator \citep[see e.g.][to name just a few]{Shen03,vanderWel14,Lange15}, the use of a radius that is related to the in situ star formation can potentially provide a radically different size evolution of these objects. Indeed, for disc galaxies with the mass of the MW, the observed size evolution using the effective radius as an indicator \citep[see e.g.][]{vanderWel14,Mowla19,Nedkova21,Kawinwanichakij21} is found to be relatively mild (with an increase in size since z$\sim$1 of only 20-30\%). This mild size evolution of disc galaxies using the effective radius is found no matter how the selection of disc galaxies was made (either by morphology or by their star-forming nature), the wavelength where r$_{e}$ was measured (usually taking the V-band restframe, approximately $\lambda$ $\sim$ 5000 $\AA$), the facility used to image the galaxies, or the different codes used to derive the galaxy structural parameters \citep[see e.g.][]{Haeussler22,Bretonniere23}. Interestingly, this reported mild  evolution provided by the use of the effective radius contrasts sharply with simple theoretical predictions based on first principles' arguments suggesting the evolution of nearly a factor of $\sim$2 of the size of disc galaxies since z$\sim$1  \citep[see e.g.][]{Mo98}

To test whether the use of the new size definition produces results more on the line of theoretical expectations, it is necessary to have very deep, high spatial resolution images. For this reason, we take advantage of the Hubble Space Telescope (HST) observations of the CANDELS fields. Deep HST images in the past have shown their power to detect and characterise low surface brightness features \citep[see e.g.][]{Trujillo05,Azzollini08,Trujillo13,Buitrago17,Borlaff19}. In order to maximise the possibility of exploring the outer parts of the galaxies with high signal-to-noise ratio and therefore identify their edges reliably, we restrict our study to only massive  (M$_{\rm stellar}$ $\gtrsim$10$^{10}$ M$_{\odot}$) disc systems. In addition, we only study  galaxies with  z $<$ 1 because of the growing surface brightness cosmological dimming -- proportional to (1+z)$^{3}$ when working with flux density (as it is the case in the AB magnitude system) for resolved objects \citep{Giavalisco1996,Law07,Ribeiro16} --.

This work is structured as follows: Section \ref{sec:data} describes the datasets we used and the selection criteria for our galaxy sample. Section \ref{sec:methodology} explains the methodology we adopted in order to derive the size of our galaxies. Section \ref{sec:results} shows the newly-derived mass-size and mass-stellar density at the edge position relations. Section \ref{sec:conclusions} reports our conclusions. The  appendices outline the complementary tests that we have carried out in order to prove the reliability of our results. Hereafter, our assumed cosmology is $\Omega_m$=0.3, $\Omega_\Lambda$=0.7 and H$_0$=70 km s$^{-1}$ Mpc$^{-1}$. We use a \citet{Chabrier03} Initial Mass Function (IMF), unless otherwise stated.  Magnitudes are provided in the AB system \citep{Oke1983}.

\section{Data and sample selection criteria}
\label{sec:data}

\begin{figure}
    \resizebox{\hsize}{!}{\includegraphics{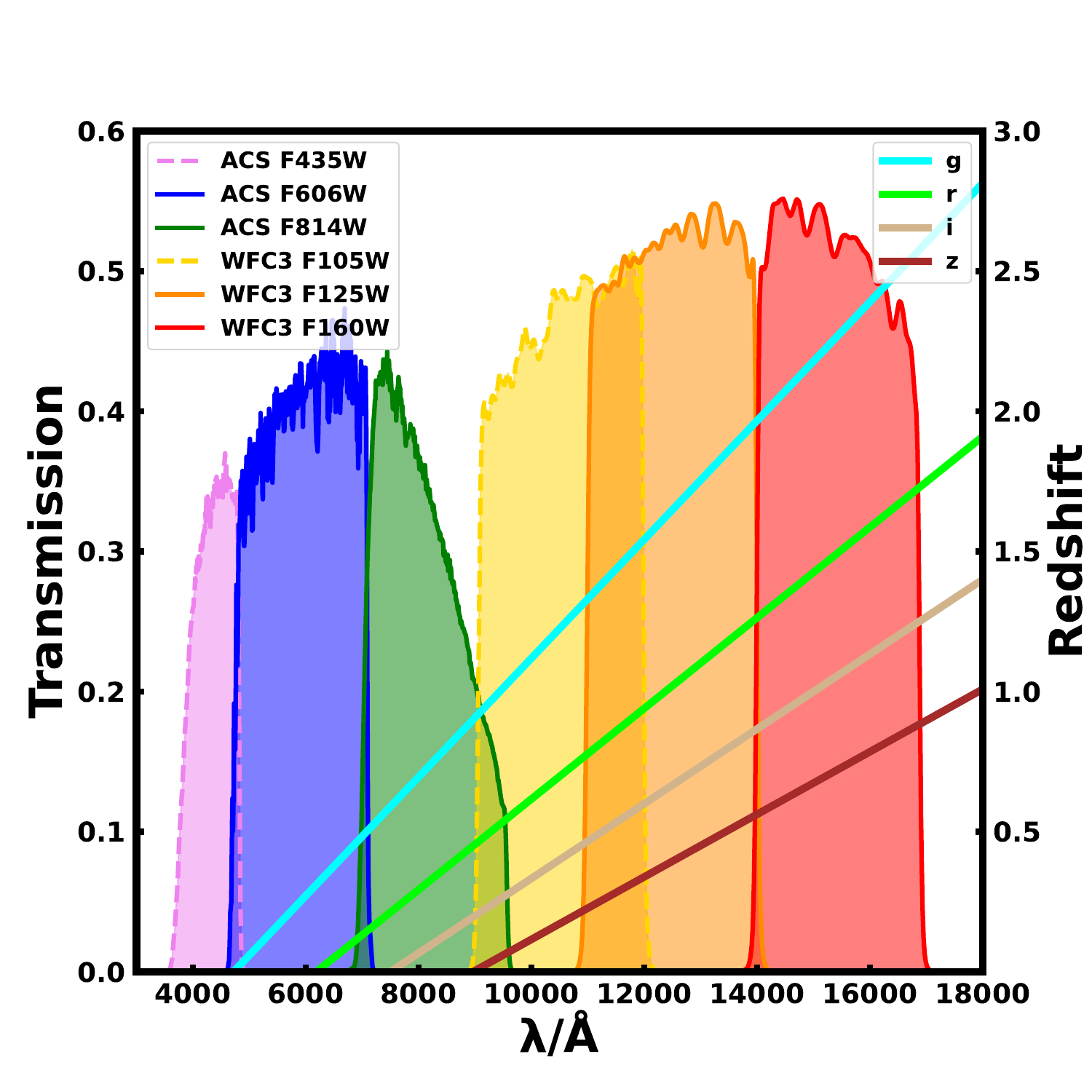}}
    \caption{Set of HST filters used in this study. The y-axis refers to the filters' transmission (left) and also the position where the restframe SDSS central wavelengths move with redshift (right). Dashed filters are only available for the GOODS-South field.}
    \label{fig:filters_and_redshifts}
\end{figure}

As our reference dataset, we use the latest available images (v1.0) in the CANDELS survey\footnote{http://arcoiris.ucolick.org/candels/} \citep{Grogin11,Koekemoer11}.
In the case of the GOODS-South field, however, we use the data from the Hubble Legacy Field\footnote{https://archive.stsci.edu/prepds/hlf/} \citep[HLF v2.0;][]{Illingworth16} which contains a larger area coverage. Additionally, for the Hubble Ultra Deep Field, we make use of its associated optical imaging \citep[]{Beckwith06} while in the near-infrared we take the ABYSS data \citep{Borlaff19}. This last survey is a new rendition of the HUDF WFC3 imaging taking special care in preserving the LSB structures\footnote{http://research.iac.es/proyecto/abyss/}. The filters we use throughout this paper are the F606W (V-band) and the F814W (I-band) from the ACS camera, and the F125W (J-band) and the F160W (H-band) from the WFC3 camera, except in the GOODS-South field where we supplement them with the ACS F435W (B-band) and the WFC3 F105W (Y-band). The set of filters used in this work are displayed in Fig. \ref{fig:filters_and_redshifts}.

Our target selection is based on the CANDELS public catalogues\footnote{https://archive.stsci.edu/prepds/candels/} \citep{Santini15,Stefanon17,Nayyeri17,Barro19}. We took the galaxy IDs, photometric masses and spectroscopic redshifts (only those flagged as the ones with top quality), discarding those objects with SExtractor stellarity $>$ 0.8. We added other high-quality spectroscopic redshifts from other sources, namely the LEGA-C DR2 redshifts with f$_{\rm use}$ = 1 \citep[][]{vanderWel16,Straatman18} and later from the DR3 data \citep{vanderWel21}, the ZCOSMOS Final Data Release \citep{Lilly09} with Confidence Class 3 and 4 and hCOSMOS \citep{Damjanov19} secure redshifts (r-value $>$ 5). With this information at hand, we kept in our sample only those galaxies with M$_{\rm stellar}$ $>$ 10$^{10}$ M$_{\odot}$ and up to a distance of z$_{\rm spec}$ $<$ 1.1, in order to optimise the signal-to-noise ratio for every object in our sample. This produces a total number of 2192 objects. We supplement these data with single S\'ersic structural parameters coming from the CANDELS public catalogs \citep{vanderWel12}. However, for well-resolved galaxies (in practice those at z $<$ 0.2), these structural parameters (namely effective radii, S\'ersic indices, axis ratios and position angles) are sometimes only representative of the luminous central galaxy parts that drive the $\chi^{2}$ galaxy fits \citep{Papaderos1996,dosReis20}. As the position angle and axis ratio of the galaxies are key to extract the surface brightness profiles, for those objects with z $<$ 0.2, we re-estimate their position angles and axis ratios by fitting an ellipse to the region with surface brightness (in the H band) between 23.5 to 24 mag/arcsec$^2$. By  doing so, we obtain a more representative axis ratio and position angle of the outer parts of the galaxies. Besides, for all galaxies, the axis ratio and position angle values were visually scrutinised and changed if necessary.

To select the disc galaxies in our sample, we gathered machine-learning derived  morphologies from \citet{Huertas-Company15}. Specifically, we select those objects that fulfil the conditions for being disc-dominated, namely those catalogued either as pure discs, bulge+disc, or irregular discs (see Section 6 of the aforementioned paper). From our initial 2192 objects, our sample was then reduced to 1442  classified as disc galaxies. We also removed: 158 interacting galaxies (as their edges are ill-defined), 101 galaxies lacking imaging in either F606W or F814W bands, 76 galaxies with either noisy imaging or located close to the survey borders, 28  extremely compact galaxies (with a visual extension less than 5 kpc) whose surface brightness profiles does not allow the identification of an obvious edge, 17 visually identified by us as elliptical galaxies, five galaxies with odd pixel values, four visually identified by us as  stars, four galaxies with non-realistic colours and one stellar spike, for a total final sample comprising 1048 galaxies.

Summarising, our galaxy sample selection criteria for our targets were:
\begin{itemize}
  \item M$_{\rm stellar}$ $>$ 10$^{10}$ M$_{\odot}$
  \item z$_{\rm spec}$ $<$ 1.1
  \item Being classified as discs in \citet{Huertas-Company15}
  \item Galaxies not being affected by any observational artefact
  \item Non-interacting galaxies
\end{itemize}

\begin{figure*}
\centering
    \includegraphics[width=\textwidth]{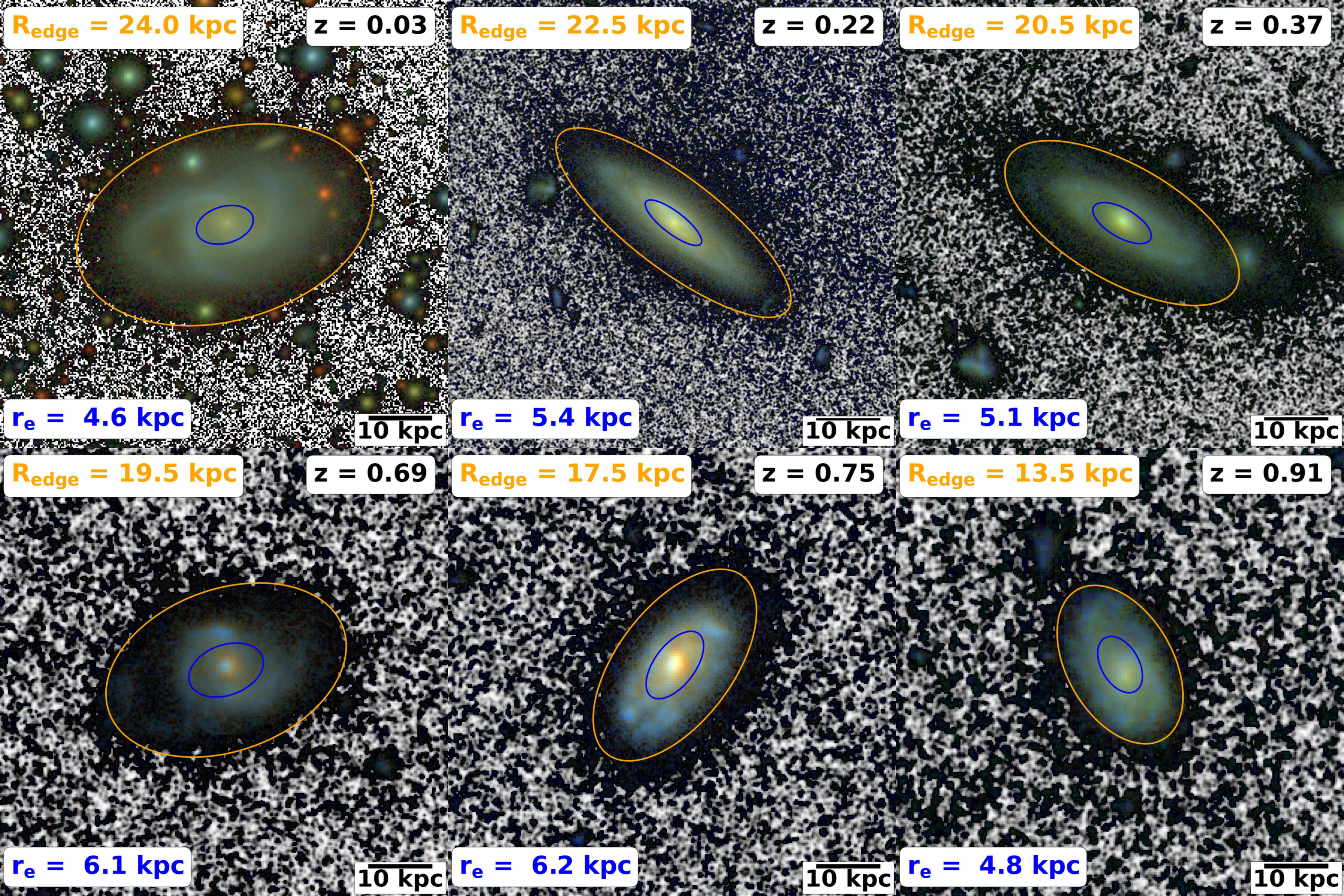} 
    \caption{MW-like (M$_{\rm stellar}$ $\sim$ 5$\times$10$^{10}$ M$_{\odot}$) galaxies  at six different  cosmic epochs in our study. Each individual stamp has a 70$\times$70 kpc size. Note that the background for each image appears in black and white in order to highlight the level at which the noise starts to dominate. The lowest redshift galaxy belongs to our Local Universe reference sample \citep{Chamba22} based on the IAC Stripe 82 Legacy Project \citep{Trujillo16}, while the rest of images are HST data. These objects were selected to have representative sizes (R$_{\rm edge}$, golden ellipses) of galaxies at those redshifts. Note how the sizes (using our in-situ star formation  criterion) decrease with redshift, while the effective radii (r$_{\rm e}$, blue ellipses) do not change appreciably.}
    \label{fig:mosaic_galaxies}
\end{figure*}

For galaxies fulfilling these conditions we created postage stamps of 150$\times$150 kpc, to a maximum of 1000$\times$1000 pixels for ACS stamps and 500$\times$500 pixels for WFC3 stamps. Some examples for MW-like galaxies in our sample at different cosmic times can be found in Figure \ref{fig:mosaic_galaxies}. From low to high redshift, their IDs are: J205432.42+000231.09 \citep[to show a nearby Universe counterpart from][]{Chamba22}, 12783\_COSMOS, 8902\_EGS, 16245\_COSMOS, 14358\_EGS and 8221\_UDS. These IDs are constructed joining the CANDELS IDs adding their respective galaxy fields.

Since our sample of galaxies is selected on the basis of spectroscopic redshifts, a potential bias in the analysis is that the sub-sample of disc galaxies with spectroscopic redshift could be biased towards objects with brighter effective surface brightness than the general disc population. What we have done to investigate this issue is to calculate the effective surface brightness in H-band of all the disc galaxies in the CANDELS sample with z$<$1 (i.e. where the redshift has been derived both photometric and spectroscopically). We have checked whether the distribution of the effective surface brightness differs depending on whether the galaxy has a spectroscopic redshift or not. We find that the distribution of effective surface brightness in the original parent sample and our spectroscopically selected sub-sample are very similar, making it unlikely that our selection criterion introduces any potential bias.

\section{Methodology}
\label{sec:methodology}

In this section, we describe the steps we have followed to determine the galaxy edges, and therefore, retrieve the size of these objects using the size definition proposed in \citet{Trujillo20} and \citet{Chamba22}.

\subsection{Observed surface brightness profiles}
\label{subsec:sb_profiles}

As mentioned in the Introduction, our proxy for the galaxy size corresponds to the radial location of the gas density threshold that enables star formation. As a consequence of this physical process, we expect this to leave its imprint on the galaxy surface brightness and radial colour profiles\footnote{For a detailed flowchart of the procedure followed in this paper, we invite the reader to take a look at Fig. 1 of \citet{Chamba22}.}. To retrieve accurate galaxy surface brightness profiles,
it is mandatory to perform a careful masking of the neighbouring objects. We created optical and near-infrared masks according to SExtractor \citep{Bertin1996} detections in the I- and H-bands, the reddest filters in the HST cameras used in this work. To be conservative, as the ACS data has large spatial resolution and therefore the deblending of the objects is more simple, we joined both masks to create the final near-infrared masks. This was done resampling the optical masks to the near-infrared spatial resolution. Our masking algorithm is an evolution of the one successfully implemented in other publications such as \citet{Trujillo07} and \citet{Buitrago08, Buitrago13}. However, the output masks have been improved for the majority of the galaxies studied to ensure that there is no obvious source of spurious light entering our analysis. This is done by manually masking very faint and small sources that could have an effect, especially at the faintest surface brightness levels.

An accurate background removal is key to determine a reliable surface brightness profile. To perform this task, we followed the strategy described in \citet{Pohlen06} and \citet{Buitrago17}, measuring the background in regions very close to the galaxies in order to be representative of the local sky background. Using as a first crude size for the galaxy the isophote at 27.5 mag/arcsec$^{2}$ in the V-band, we estimate the local background using pixels located in elliptical apertures with semi-major axes between 2 to 3 times such rough size. That is done in every filter, and then we subtracted the mean background flux within that annulus. 

The determination of the galaxy size according to the definition used in this work relies on the determination of a sharp change (i.e. a truncation) on the surface brightness (stellar surface mass density) profiles of the galaxies. Traditionally, surface brightness profiles have been derived by azimuthally averaging the galaxy flux in annulli at progressive further galactocentric distances. We call this technique the 'ellipse method'. It has the advantage of increasing the signal-to-noise of the profiles but, on the other hand, this method is also prone to blur the truncations due to its mixing of light from combining potentially different radial locations of the truncations depending on the azimuthal angle. This is the reason behind computing the surface brightness profiles in two manners in the present work, using both this ellipse method and along the semi-major axis of the galaxies (calling this latter one the 'slit method'). For the first technique, we derived the galaxy fluxes using resistant mean determinations (i.e. removing 3$\sigma$ outliers) in one-kpc-wide elliptical annulli. The centre, axis ratio and position angle of the elliptical annulli were fixed to the values describing the light distribution of the galaxies retrieved in the H-band images (see Section \ref{sec:data}). H-band images are used to avoid dust obscuration effects and to better track the galaxy stellar mass. On the other hand, we used the slit method to trace the surface brightness profiles of the galaxies along their semi-major axis. This is the strategy usually performed for edge-on disc galaxies \citep[e.g.][]{Martin-Navarro12}. The disadvantage of the slit method with respect to the ellipse method is that the signal-to-noise is poorer due to the lower number of pixels used. To minimise as much as possible this issue, when creating the semi-major axis profiles, our slit apertures had a width of 3 kpc and we summed both sides of the galaxy. This method is superior for dealing with highly inclined galaxies and also for finding sharp drops in the disc light, as only pixels in a given direction contribute to the final statistics. We have studied the location of the edge of galaxies using both the ellipse method and the slit method. We find that the location of the edge can vary by about 10\% from one method to the other. Larger deviations of this variation can be expected if the outer parts of the galaxies are irregular.

\subsection{Sloan-restframe equivalent profiles}
\label{subsec:sdss}

To derive the  stellar mass density profiles, we converted the observed surface brightness profiles into their equivalent in Sloan restframe filters. To that end, we took advantage of the already existing recipes to derive mass-to-light ratios from Sloan colours \citep[see e.g.][]{Bell03,Roediger15}. In order to perform this task correctly, we corrected the observed surface brightness profiles using the following steps: 

\begin{itemize}
  \item Galactic extinction correction. To do this, we used the values provided by the NASA NED Extinction Calculator\footnote{https.//ned.ipac.caltech.edu/extinction\_calculator} --
  \begin{equation}
  \mu_{corr,1} = \mu_{obs} - \mu_{extinction}
  \end{equation}
  with $\mu_{corr,1}$ and $\mu_{obs}$ being the corrected and observed surface brightness profiles, respectively.
  \item Cosmological surface brightness dimming correction. We use (1+z)$^3$ as we are working with flux densities as explained before:
  \begin{equation}
  \mu_{corr,2} = \mu_{corr,1} - 7.5\times \rm log_{10}(1+z)
  \end{equation}
  \item Disc inclination correction. For this last step, we used the expression
  \begin{equation}
  \mu_{corr,3} = \mu_{corr,2} + \sum^{4}_{j=0} \alpha_{j} \left( \sfrac{b}{a} \right)^{j}
  \end{equation}
  assuming $z_{0}/h$ = 0.12  \citep[for the determination of the $\alpha_{j}$ coefficients see Section 5.2 in][]{Trujillo20}, where z$_{0}$ indicates the disc's model scale height and $h$ its scale length--
\end{itemize}

 We do not correct our surface brightness profiles for internal dust extinction. While this effect is potentially moderate for low-inclination galaxies, it could be relevant for the galaxies in our sample with high inclination.

The Sloan restframe profiles are calculated following the procedure in \citet{Buitrago17}. We interpolated the observed surface brightness profiles linearly between contiguous observed bands in order to obtain the restframe Sloan magnitudes. From them, we also derived radial colour profiles. To further improve our results, we also checked whether the derived Sloan colour values were consistent with the predictions of the E-MILES library\footnote{http://research.iac.es/proyecto/miles/pages/photometric-predictions-based-on-e-miles-seds.php} \citep{Vazdekis12,Ricciardelli12,Vazdekis16}. We explicitly check that the Sloan restframe colours were consistent with stellar population ages smaller than the age of the Universe at each galaxy's redshift. We use the predictions provided by the Padova+00 isochrones \citep{Girardi00} and a \citet{Chabrier03} IMF for any given metallicity. If the Sloan colours are compatible with ages older than the Universe at that cosmic time, we marked these colour points as unreliable. We also mark as unreliable  those points in the radial colour profiles with errors greater than 0.2 magnitudes.

\subsection{Stellar mass density profiles}
\label{subsec:trunc_position}

We created the stellar mass surface density profiles following the scheme described in \citet{Bakos08} that utilises the following expressions:

\begin{equation}
\rm log_{10} \Sigma_{\star} = log_{10} \left( \sfrac{M}{L} \right)_{\lambda} - 0.4\times (\mu_{\lambda} - m_{abs,\odot,\lambda}) + 8.629
\end{equation}

where m$_{abs,\odot,\lambda}$ is the absolute magnitude of the Sun at the filter\footnote{Taken from http://mips.as.arizona.edu/~cnaw/sun.html} corresponding to the wavelength $\lambda$ and $\mu_{\lambda}$ is the corrected surface brightness profile using that filter. The final result is given in units of M$_{\odot}$/pc$^{2}$. The mass-to-light ratio at the filter corresponding to $\lambda$ is calculated as follows:

\begin{equation}
\rm log_{10} \left( \sfrac{M}{L} \right)_{\lambda} = a_{\lambda} + (b_{\lambda} \times \rm colour)
\end{equation}

where the coefficients $a_{\lambda}$ and $b_{\lambda}$ are tabulated in \citet{Roediger15}, assuming a \citet{Chabrier03} IMF. We obtained all possible combinations of base profile ($g$, $r$, $i$, $z$) and colour ($g-r$, $g-i$, $g-z$, $r-i$, $r-z$, $i-z$) to create as many as possible stellar mass density profiles. This gives us a total of 24 different stellar mass density profiles. The final stellar mass density profiles we use in this work are the resistant mean combination of each individual stellar mass profile generated above. The estimation of the errors is done by taking into account the individual errors for each individual mass profile (which is basically given by the error in the colour used to compute it) and the intrinsic dispersion of the 24 independent mass profiles. The latter arise from the uncertainty of the \citet{Roediger15} method and reflects the error in the M/L determination. Both errors are summed in quadrature. As for the Galactic extinction and inclination errors, they are not taken into account since both corrections are systematic and identical for all profiles.

As a consistency test, we derived the total  stellar masses of our galaxies using the derived stellar mass density profiles. For galaxies displaying low inclinations (axis ratio $>$ 0.3), the total stellar masses were derived using the stellar mass density profiles derived from the ellipse method. For galaxies displaying high inclinations (axis ratio $<$ 0.3), the  total stellar masses were derived using the stellar mass density profiles derived from the slit method. It is important to note that this is done to better reflect the observed symmetry of the object under study and to minimise the potential effect of the thickness of the disc in our total stellar mass estimations. The integration of the radial profiles to get the total stellar mass cover from zero to the last radial point consider reliable (see Section \ref{subsec:sdss}). This last restriction includes situations when the stellar mass density profile started to rise as a consequence of being artificially affected by the light of a neighbour source.

\begin{figure}
    \resizebox{\hsize}{!}{\includegraphics{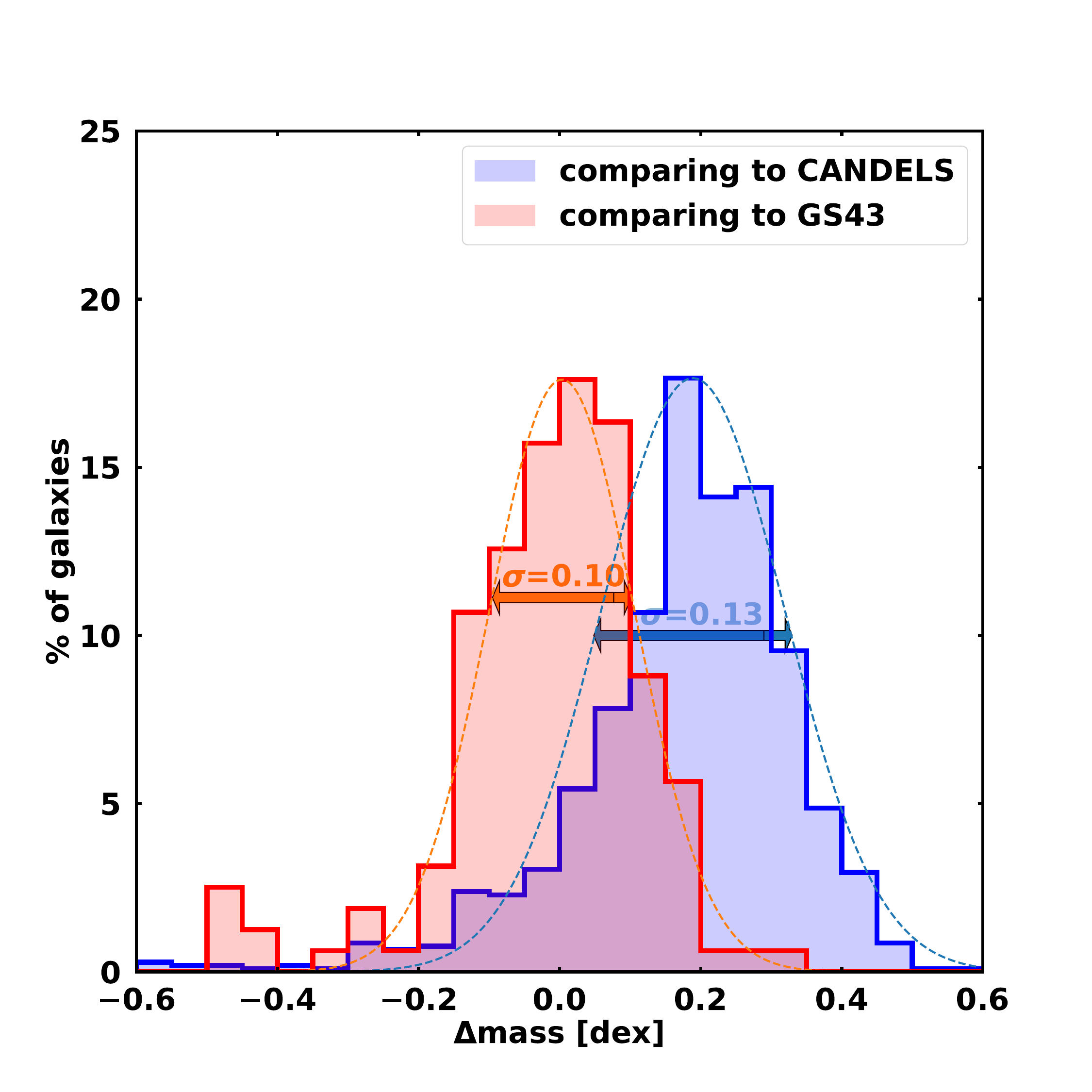}}
    \caption{Comparison between our photometrically derived stellar masses and those in the CANDELS and ASTRODEEP GS43 catalogues ($\Delta$mass = log(M$_{\rm ours}$)-log(M$_{\rm catalog}$)). The dashed lines represent the best Gaussian fits (their scatter is also displayed). The arrows with the sigma values are displaced vertically for the blue distribution for the sake of clarity.}
    \label{fig:masses}
\end{figure}

Our integrated stellar masses are compared to the CANDELS-derived masses (see Section \ref{sec:data}) and the recent ASTRODEEP GS43 catalog \citep{Merlin21}. This last one is a state-of-the-art catalog containing consistent photometry for 43 medium- and broad-filters available in the GOODS-South field. There are a total of 159 galaxies overlapping with our sample in this case. The outcome of this comparison is shown in Fig. \ref{fig:masses}. While both CANDELS and GS43 masses display a similar scatter in $\Delta$mass (=log(M$_{\rm ours}$)-log(M$_{\rm catalog}$)) with respect to our determinations, 0.13 and 0.10 dex for CANDELS and GS43 respectively, there is an offset with respect to CANDELS masses (0.19 dex). This is not found in the more comprehensive ASTRODEEP GS43 catalog (0.003 dex). This test gives us confidence on our stellar mass density profiles derivation. Therefore, on what follows,  we adopt the total stellar masses derived from the integration of our stellar mass density profiles. 

\subsection{Deriving the size of the galaxies}
\label{subsec:trunc_position}

We use the radial position of the truncations in our profiles as a proxy to derive the physically motivated size radius. Following Fig. 1 in \citet{Chamba22}, the truncation positions (R$_{\rm edge}$) are identified as sharp changes in the galaxy surface brightness, colour and/or stellar mass density profiles in the outermost galaxy regions. On what follows, the truncation feature is dubbed as edge to reinforce the idea that our proxy traces the end of the in-situ star forming disc. When the edges are not obvious on the stellar surface mass density and/or surface brightness profiles, the radial colour profiles are particularly useful to retrieve the radial position of the edge. In fact, \citet{Pfeffer22} and \citet{Chamba22} show that the rising of the characteristic U-shape in their colour profiles \citep{Azzollini08,Bakos12,Watkins16,Watkins19,Martinez-Lombilla19} is connected with a significant drop in the star formation of the galaxy. At that radial position, the disc light diminishes while the potential emergence of a stellar halo gains in importance creating a redder colour. Therefore, using that feature can help us to easily identify the location of the edge of the disc.

Quantitatively, the process we have followed to identify galaxy edges is as follows. Based on the work done in the present-day Universe on Milky Way-like edge-on galaxies such as NGC4565 and NGC5907 \citep[see][]{Martinez-Lombilla19}, we expect the change on slope of the stellar surface mass density profiles to be around 2-3 (i.e. h1/h2$\sim$2-3, where h1 is the scale length of the disc before the truncation and h2 is the scale length after the truncation). For galaxies with lower inclinations, the change in slope at the edge is expected to be milder for a number of reasons. First, the contrast (and therefore the S/N) between the inner disc and the outer region of the galaxy decreases due to the line-of-sight integration. Second, the use of ellipse profiles (instead of slits) averages light from different angles of the galaxy, and if the edge is not at exactly the same radial distance in all directions, this will also blur the edge detection. We have simulated \citep[using IMFIT;][]{Erwin15} a disc galaxy with different inclinations to see what change in the h1/h2 ratio we would expect. The result of this analysis is that even for a perfectly symmetric disc, if we start with a value of h1/h2=3 for an edge-on configuration, for a lower inclination we measured values with h1/h2$\sim$1.5. In real galaxies, taking into account the effects of the point spread function (PSF), deviations from symmetry, etc., 1$<$h1/h2$<$1.5 is then expected.

From our initial sample of 1048 disc galaxies, about 60\% of them (i.e. 634) show a change in the slope in the outer parts of the stellar surface mass density profiles that is compatible with the expectation (i.e. h1/h2$>$1). In cases where h1/h2$\sim$1 (within the errors in their determination), the use of the stellar surface mass density profiles is not sufficient for our purposes to detect an edge with some confidence. For this reason, our next step is to use the surface brightness profiles in the observed HST bands V, I, J and H and to re-estimate h1/h2 around the potential location of the edge. If h1/h2 is greater than 1 then we use such a value as the location of the edge. This gives a further 8 galaxies with an identified edge.

If the value of h1/h2 is still not greater than 1 with sufficient confidence, then we examine the shape of the rest-frame colours (mainly g-r) to obtain an estimate of the edge position. In this case, we take advantage of the well-known U-shape of the colour profiles for disc galaxies (see e.g. Bakos et al. 2008). This includes a further 287 galaxies with edges identified. Only 119 galaxies (11\% of our sample) have no clear edges. In these cases, we identify the edge as the visual limit to which we can see the galaxies in our images. Finally, once the location of the edge (R$_{\rm edge}$) is found, the stellar mass surface density at that position ($\Sigma$$_{\rm edge}$) is what we assign as the stellar mass surface density at the edge. In Appendix \ref{app:jwst}, we show some examples of the expected variation in the determination of the R$_{\rm edge}$ using the different methods. In general, we find good agreement (with uncertainties on the R$_{\rm edge}$ position within 1 kpc uncertainty). In addition, we probe the robustness of the size determination using recently released very deep and superior spatial resolution JWST data \citep{2023arXiv230602465E,2023arXiv230602466R}.

To test the accuracy of our approach, the authors (FB, IT) independently carried out the determination of galaxy sizes (R$_{\rm edge}$) by visually comparing all the profiles and images for the galaxies in our sample. To be sure that the determination of the edge in our profiles is not an artefact of the noise of our images, we calculated the limiting surface brightness of each our images. To make this estimate, we randomly selected pixels on the masked images and calculated the scatter of the pixel distributions in apertures corresponding to 1$\times$1 arcsec. Our limiting surface brightness corresponds to a 3$\sigma$ fluctuation of the background noise on these areas. A detailed description of the calculation of these limits is given in \citet{Roman20}. While the background pixel noise is a simple approximation to characterise the true surface brightness limit of the image, especially in the presence of large scale light gradients that may have been left over during data reduction, in the particular case of the galaxies we are working with, which are significantly smaller than the size of the HST detectors, we believe that our local characterisation of the background noise is a fair representation of the local surface brightness limits around such galaxies.

The edge feature and, therefore, the size determination is found in all (except for 11\% of the cases, see above) to be located at radial distance where the observed surface brightness profiles are brighter than the limiting surface brightness  for each photometric band. Once the radial position of the edge is found, its correspondent value in the stellar mass density profile determines the stellar mass density at the edge ($\rm \Sigma_{\rm edge}$) parameter, using the ellipse method for galaxies with axis ratios ar $>$ 0.3 and the slit method for ar $<$ 0.3.

\begin{figure*}
\centering
    \includegraphics[width=0.89\textwidth]{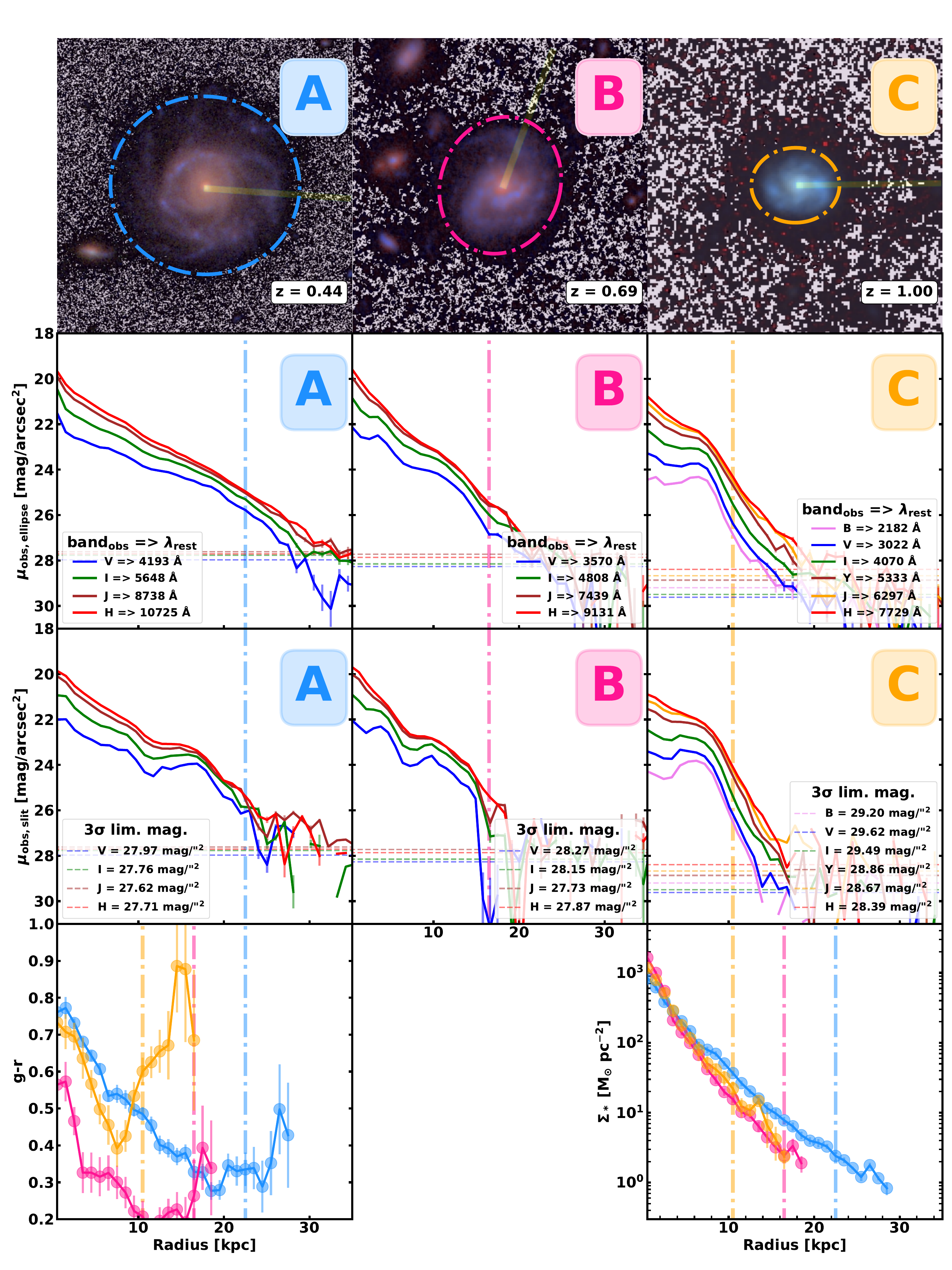}
    \caption{Summary of our analysis for three MW-like galaxies at different redshifts (from left to right): A) 9217\_COSMOS at $z$ = 0.44 in blue, B) 20537\_EGS at $z$ = 0.69 in magenta, and C) 12887\_GOODSS at $z$ = 0.997 in dark yellow. The first row displays their colour RGB images with their sizes (R$_{\rm edge}$) highlighted by dot-dashed lines whose positions are also marked in the plots below. The orientation of the 'slit' we used to extract the slit profiles plotted in the panels of the third row is shown in yellow. The second row shows the surface brightness profiles in V-, I-, J- and H-band using azimuthally averaged profiles ('ellipse method') while the horizontal lines denote the images' observed 3$\sigma$ limiting surface brightness magnitudes in 1$\times$1 arcsec$^{2}$ boxes. The legend gives the equivalent restframe central wavelength for each observed filter. The third row is similar to the second, but this time following the semi-major axis ('slit method') profiles. The observed limiting surface brightness is written explicitly this time in the legend. For the fourth row, the bottom left plot shows the $g-r$ colour profiles and the bottom right the stellar surface mass density profiles.}.
    \label{fig:galaxy_profiles}
\end{figure*}

We display in  Figure \ref{fig:galaxy_profiles}, some examples of our derived profiles/quantities in order to determine the size of three MW-like galaxies (columns). There, we show the following panels (rows from top to bottom) --with the in situ star formation radius (R$_{\rm edge}$) appearing as a dash-dotted vertical line with a different colour for each galaxy--:

\begin{itemize}
  \item The \textit{first row} shows colour composite images for the galaxies A) 9217\_COSMOS, B) 20537\_EGS and C) 12887\_GOODSS. These RGB images were generated, together with the rest of colour images in the paper, using the algorithm \textit{rgb-asinh} in the \texttt{gnuastro}\footnote{https://www.gnu.org/software/gnuastro/} software package \citep{Akhlaghi15,Akhlaghi19}. The edge position appears as a dot-dashed line, the same that also marks its position in the panels below. The slit orientation we utilised for deriving the ``slit method" profiles (see third item in this list) is shown in yellow on top of the galaxy images.
  \item The \textit{second row} shows the observed surface brightness profiles (i.e. using the ``ellipse method"). The legend indicates the equivalent restframe wavelength that each filter's central wavelength is probing. The dashed horizontal lines in the bottom are the observed 3$\sigma$ limiting surface brightness magnitude derived using 1$\times$1 arcsec$^2$ apertures for each image. As expected for a radius related with star formation, edges are more clearly detected in the bluer bands. One can also easily identify the bulge and the spiral structures in these profiles.
  \item The \textit{third row} shows the observed surface brightness profiles along the semi-major axis (i.e. using the ``slit method"). The legend indicates again the observed 3$\sigma$ surface brightness limits. Note that, as expected the profiles are noisier with respect to the azimuthally averaged ones. However, the edges are more obvious as the irregular shape of the light distribution of these galaxies do not blur the edge features resulting from the ellipse method.
  \item The \textit{fourth row, bottom left position}, shows the restframe $g-r$ colour profiles. The individual profiles were corrected by galaxy inclination, Galactic extinction and cosmological dimming as indicated in Section \ref{subsec:sdss}. While we have created all colours possible with the derived Sloan-equivalent profiles, we chose to show only $g-r$ for the sake of clarity. They display their characteristic U-shape. When the stellar mass and surface brightness profiles do not clearly show the edge, colour profiles are key to elucidate its radial position.
  \item The \textit{fourth row, bottom right position}, shows the averaged stellar surface mass density profiles based on the ellipse method. Using all possible combinations of base profiles and colours, we created resistant mean profiles. The stellar surface mass density ($\rm \Sigma_{\rm edge}$) at the edge position is derived from these profiles.
\end{itemize}

The main parameters of our sample --coordinates, redshift, masses, sizes (R$_{\rm edge}$) and the stellar surface mass densities at the edge positions ($\rm \Sigma_{\rm edge}$)-- could be found in Table \ref{tab:sample_params}. Its full version is included in the digital version of this article, where we also provide the surface brightness (both the observed and the Sloan rest-frame equivalent) and the stellar surface mass density profiles from our sample of galaxies. The  latter (i.e. the Sloan rest-frame equivalent and the stellar surface mass density profiles) are corrected for Galactic extinction and inclination.

\begin{table*}
\centering
\caption{Galaxy sample main parameters. The full table is available at the CDS.}
\begin{tabular}{cccccccc}
\hline \hline
Galaxy ID & Field & RA & DEC & z & Mass & R$_{\rm edge}$ & $\rm \Sigma_{\rm edge}$ \\
 & & [J2000] & [J2000] & & [$\times10^{10}$M$_{\odot}$] & [kpc] & [M$_{\odot}/\rm pc^{2}$] \\
\hline
270 & COSMOS & 150.0824879 & 2.1821822 & 0.680 & 6.33 $\pm$ 2.15 & 13.5 $\pm$ 1.0 & 3.77 $\pm$ 0.13 \\
283 & COSMOS & 150.0723594 & 2.1830004 & 0.987 & 1.32 $\pm$ 0.45 & 8.5 $\pm$ 1.0  & 10.64 $\pm$ 1.89\\
886 & COSMOS & 150.1188538 & 2.1899541 & 0.346 & 1.07 $\pm$ 0.36 & 10.5 $\pm$ 1.0 & 3.52 $\pm$ 0.92 \\
... & ...    & ...         & ...       & ...   & ...   & ...  & ...   \\  
\hline
\end{tabular}
\label{tab:sample_params}
\end{table*}

\section{Results}
\label{sec:results}

\subsection{The size-mass relation}
\label{subsec:mass-size}

\begin{figure*}
\centering
    \includegraphics[width=\textwidth]{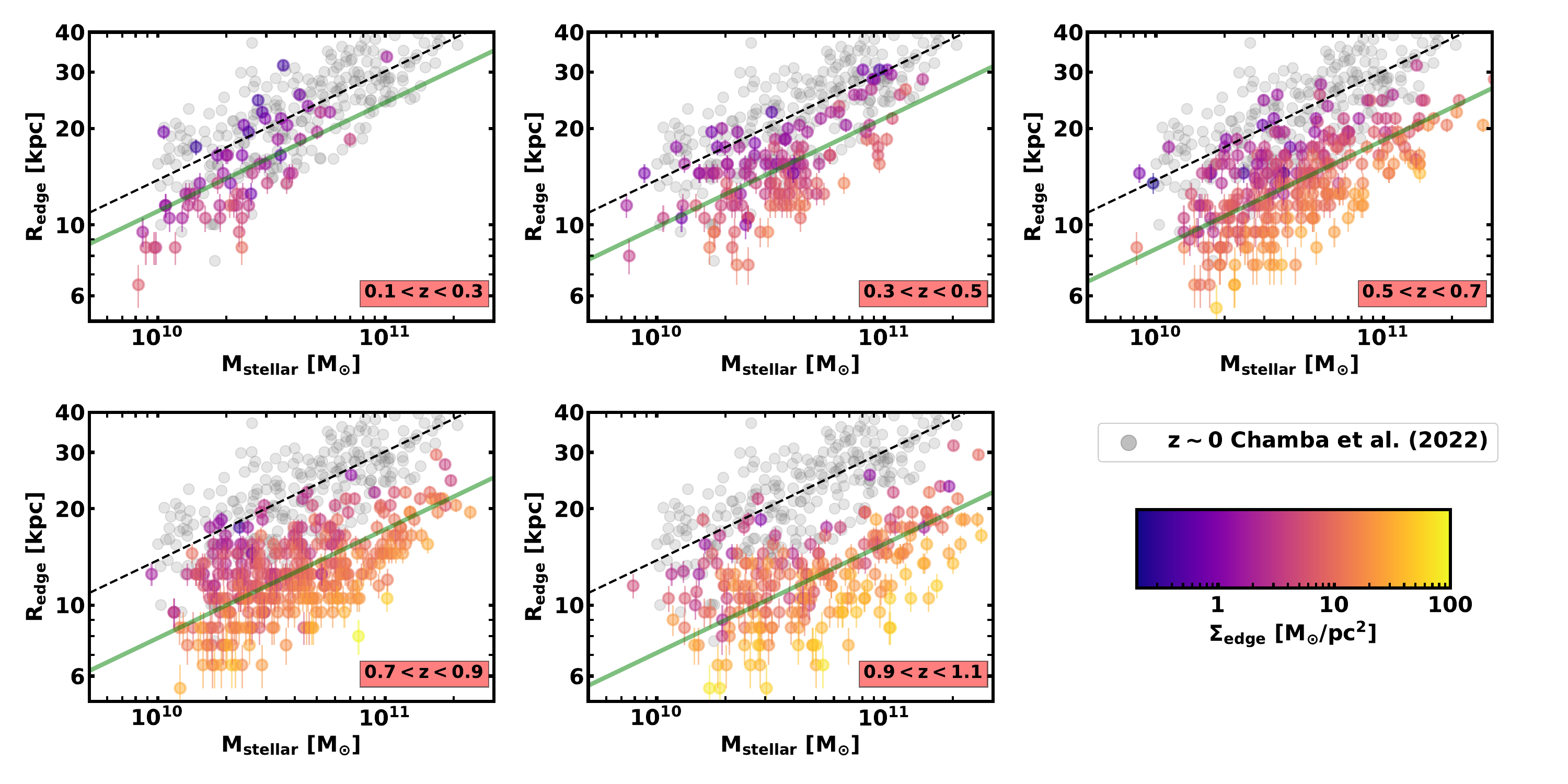}
    \caption{Galaxy size (using as a proxy the expected radial location of the end of the in situ star formation, i.e. R$_{\rm edge}$) versus galaxy stellar mass for our galaxy sample. The black dashed line represents the linear regression (in log-log plane) to the local values (in grey, visible at all redshifts) using the \citet{Chamba22} sample and the green solid line depicts the linear regression at each redshift (fixing the slope to the local value). The galaxies in our sample are colour coded based on their stellar surface mass density  at the edge position ($\rm \Sigma_{\rm edge}$). Individual error bars in our data points come from our spatial sampling when obtaining the surface brightness profiles (i.e. 1 kpc). A progressive departure from the nearby Universe values towards smaller sizes at higher redshift is noticeable. At a fixed stellar mass, more compact galaxies have larger $\rm \Sigma_{\rm edge}$.}
    \label{fig:trunc_mass_rel_colored_sigma}
\end{figure*}

Figure \ref{fig:trunc_mass_rel_colored_sigma} shows the size of the disc galaxies (using R$_{\rm edge}$ as the size proxy due to their connection with the gas density threshold enabling star formation) versus the galaxy stellar mass for five different redshift bins. 

Our local reference are the disc galaxies from the sample by \citet{Chamba22} whose sizes are determined in a similar way than ours. Although the \citet{Chamba22} sample covers all morphological types, we only show the local objects fulfilling a similar selection criteria as ours. This means taking galaxies with a visually clear disc component, that is, we select those local galaxies with Hubble type $>$ -2 (i.e. from S0 galaxies to later types). This local sample is displayed with grey colour data points in Fig. \ref{fig:trunc_mass_rel_colored_sigma}, while the galaxies analysed in this work are coloured according to their stellar surface mass density at the radial position of the edge (i.e. $\rm \Sigma_{\rm edge}$). R$_{\rm edge}$ error bars come from our spatial sampling when obtaining the surface brightness profiles (i.e. 1 kpc). The more massive disc galaxies are larger in general at any redshift. There is also a progressive decrease of the size of the galaxies with redshift. This decrease of the size of the objects is independent of their stellar mass. Similar trends are also found in the evolution of the size-mass relation  when using the effective radius as a  proxy for the galaxy size. 

To quantify the  global evolution of the size-mass relation of our galaxies, we have fitted the size-mass relation with a power-law and measure the change on the y-intercept with redshift. A visual inspection of the evolution of the relationship suggests (as a first approximation) that the slope of the size-mass relation does not change much with redshift. Therefore, to simplify the analysis, we fix the slope to the one found in the local sample (i.e. b$_R$=0.34). The fit to the local sample is always shown as a  black dashed line, while the fit at each
redshift range appears as a green line. The slope and the y-intercept values can be found in Table \ref{tab:trunc_fits} with the R subscript. The errors for the y-intercept parameters, as well as the observed scatter values for the size-mass relation at the different redshifts, are derived from bootstrapping half of the sample using 10$^{4}$ different realisations. One can see that the scatters are fairly constant ($\sim$0.1 dex) at all redshift bins.

The observed scatter for the size-mass relation at all redshifts is the result of both the intrinsic scatter of the relation plus the observational uncertainties at measuring both the size and stellar mass of the galaxies. To quantify which fraction of the observed scatter is due to uncertainties at the stellar mass determination, we follow the strategy depicted in \citet{Trujillo20}. We place our galaxies in the fitted size-mass relation at a given redshift (i.e. at each stellar mass, we associate a size corresponding to the fitting line of that size-mass relation). Then, we randomly generate artificial samples by allowing a stellar mass variation following a Gaussian distribution with a scatter of 0.13 dex (i.e. the error in our stellar mass determination according to the comparison with CANDELS\footnote{Although the comparison with the GS43 sample suggests a lower error on the stellar mass determination ($\sigma$=0.10 dex), we have decided to use the most conservative value of 0.13 dex found at comparing with the CANDELS sample.}). Then we measure the scatter of these artificial samples. The scatter associated with the uncertainty in the stellar mass (for each redshift)  appears in Table \ref{tab:trunc_fits_intrinsic} column $\sigma_{\rm R,mass}$. Of course, this uncertainty in the stellar mass has also an impact in the $\rm \Sigma_{\rm edge}$-mass relation and it appears in Table \ref{tab:trunc_fits_intrinsic} column $\sigma_{\rm \Sigma,mass}$. We add the values for the local sample for completeness. The other source of observational scatter in the size-mass relation is the uncertainty at measuring the position of the edge of the galaxies. This effect is not straightforward to model as it depends on the increasingly lower spatial resolution at higher redshift and the spatial binning we have used (1 kpc) to increase the signal-to-noise of the surface brightness profiles. Again, to have a rough estimation of this error, as we did for measuring the observed scatter due to the stellar mass uncertainty, we place our galaxies in the fitted size-mass relation at a given redshift. Then, we randomly generate artificial samples by allowing a size (R$_{\rm edge}$) variation following a Gaussian distribution with a scatter of 1 kpc because our inherent spatial discretisation. The contribution to the observed scatter in the R$_{\rm edge}$-mass and $\rm \Sigma_{\rm edge}$-mass relation can also be found in Table \ref{tab:trunc_fits_intrinsic}. There, the intrinsic scatter after correcting quadratically by both the size and mass uncertainty is given by columns $\sigma_{\rm R,int}$ and $\sigma_{\rm \Sigma,int}$. Taking into account that we have not modelled all the possible effects that can affect the location of R$_{\rm edge}$, as the uncertainty in the background removal of the images, the reported intrinsic scatter of our relationships should be considered as upper limits \citep[see a comprehensive discussion in ][]{Stone21}.

\subsection{The mass-$\rm \Sigma_{\rm edge}$ relation}
\label{subsec:mass-sigma}

\begin{figure*}
\centering
    \includegraphics[width=\textwidth]{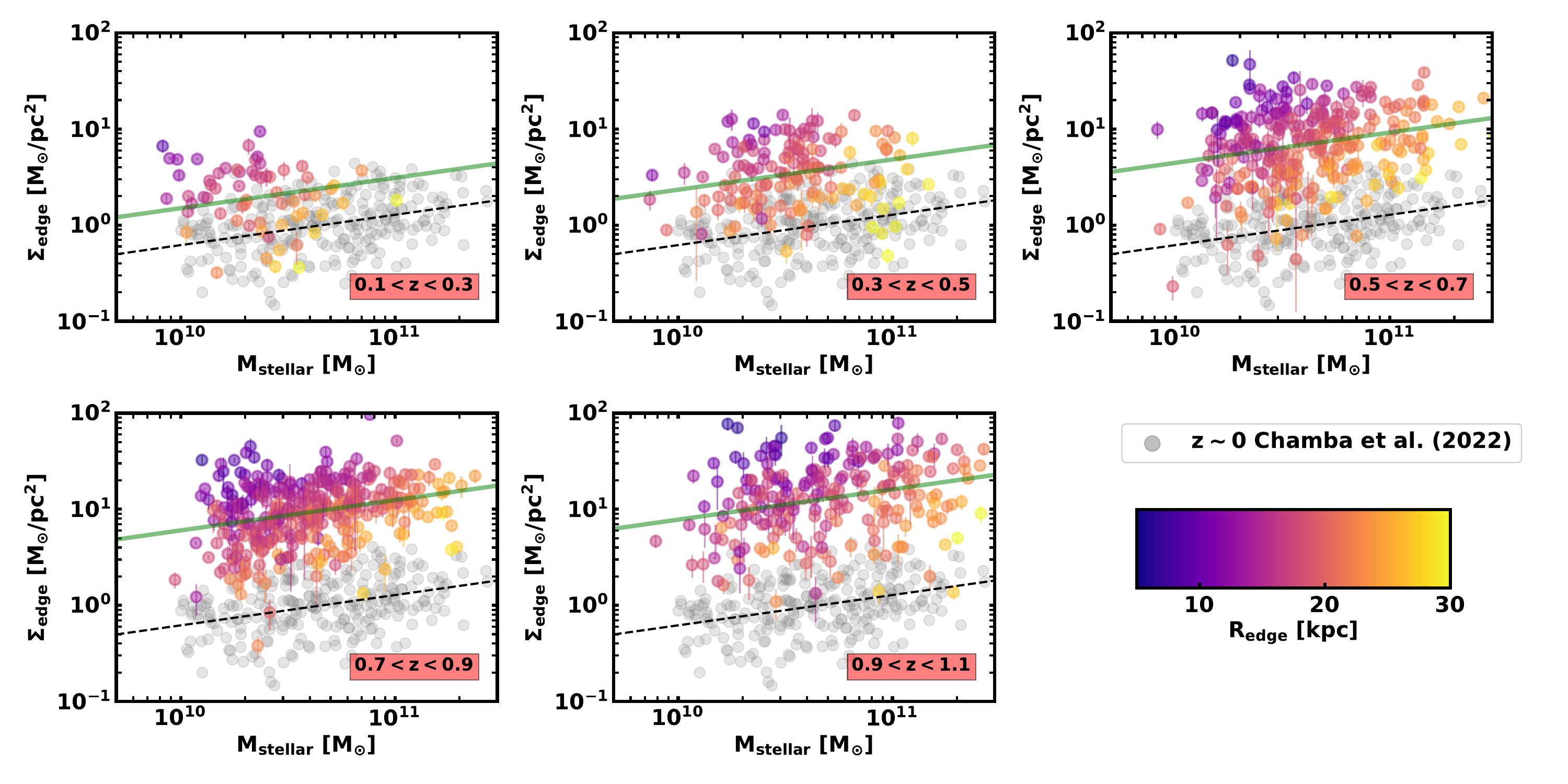}
    \caption{Stellar surface mass densities ($\rm \Sigma_{\rm edge}$) at the radial position of the edge versus galaxy stellar mass for our galaxy sample. The black dashed line represents the linear regression to the local values from \citet{Chamba22} (in grey, visible at all redshifts) and the green solid line depicts the linear regression fitting at each redshift. The data points are coloured according to their size (R$_{\rm edge}$) and the $\rm \Sigma_{\rm edge}$ error bars come from the error bars in the  stellar mass density profiles at the R$_{\rm edge}$ position. A progressive departure from the nearby Universe values towards higher densities is appreciable. The data points are colour coded according their galaxy size. At a fixed stellar mass, those galaxies with the larger stellar surface mass densities are those smaller in size.}
    \label{fig:sigma_mass_rel_colored_trunc}
\end{figure*}

Figure \ref{fig:sigma_mass_rel_colored_trunc} shows the stellar mass surface density at the edge of the galaxy ($\rm \Sigma_{\rm edge}$) versus the galaxy stellar mass. We utilise the colour coding in a similar way as before: grey
for the local sample, and the colours for showing the radial location of the edge (i.e. the galaxy size). $\rm \Sigma_{\rm edge}$ error bars come from the error bars in the  stellar surface mass density profiles at the R$_{\rm edge}$ position. As it was shown previously for the size-mass relation, the $\rm \Sigma_{\rm edge}$-mass relation of the disc galaxies is well described by a power law at all redshift bins. The least-squares fit for the \citet{Chamba22} sample is the black dashed line, while the green one denotes the fits for the higher-z samples with the slope fixed to the local value (b$_{\Sigma}$=0.32). The fits can also be found in Table \ref{tab:trunc_fits} --those with the $\rm \Sigma$ subscript--, and are computed in the same fashion as those for Fig. \ref{fig:trunc_mass_rel_colored_sigma}.

The evolution of $\rm \Sigma_{\rm edge}$ with cosmic time is as follows: first, more massive galaxies present larger $\rm \Sigma_{\rm edge}$ values and, second, there is a progressive departure from the nearby Universe values towards higher densities as the redshift increases. This last trend is even more noticeable than for Figure \ref{fig:trunc_mass_rel_colored_sigma} since the shift is larger than the one for galaxy sizes. In addition, the colour gradient in our plots shows that at fixed stellar mass,  the larger the stellar mass density at the radial position of the edge, the smaller the galaxy size. The observed and intrinsic scatters ($\sigma_{\rm \Sigma,mass}$ and $\sigma_{\rm \Sigma,int}$) are a factor of $\sim$3 larger than the ones measured in the size-mass relation. As it is the case with the size-mass relation, the observed scatters are also fairly constant with redshift. 

Because of the cosmological dimming one would expect that there would be an observational bias against observing the galaxy edges as the redshift of our galaxies grows. That should produce an artificial trend favouring only the detection of galaxy edges that are located at higher stellar densities (i.e. shorter radial distances) as the redshift increases. To explore whether this is affecting our results, we have estimated what is the minimum stellar mass density that we can reliably explore at the different redshifts. By reliable here we mean \citep[based on our experience with galaxies in the local Universe, see e.g.][]{Chamba22} that the surface brightness at the edge of the galaxy is about 2 magnitudes brighter than the limiting surface brightness of the data used to identify the boundaries of the galaxy. We have used two different approaches, one where we assume the age of the stellar population at the galaxy edge is fixed and the same at all redshift, and another where the age of the stellar population is becoming older as the cosmic time progresses. We start both approaches by assuming an age of 2 Gyr and a metallicity of [M/H] = -0.4 at z=1. The reason for taking a subsolar metallicity is because we are representing the stellar populations at the edge of the galaxies. Taking into account our observed surface brightness limits at each redshift, if the stellar populations at all redshift were 2 Gyr (because there is a continuous star formation rejuvenating the stellar population), then we should be able to detect the edge of the galaxies up to stellar mass density of $\sim$1.3 M$_{\odot}$/pc$^{2}$ at z=1 and $\sim$0.6 M$_{\odot}$/pc$^{2}$ at z=0.5. This is  below the values we observe which at those redshifts are larger than 2 M$_{\odot}$/pc$^{2}$ (see Table \ref{tab:mw-like}).  In the case where the galaxy does not form more stars at redshift lower than 1, then our detection limits would be very similar with the limit for detecting stellar mass density at z=0.5 being $\sim$0.9 M$_{\odot}$/pc$^{2}$.

In addition to the previous analysis, we have conducted a large number of tests to explore the robustness of the evolution of the size-mass and stellar surface mass density-mass relations we have found. These tests are all illustrated with figures in the Appendices. Here we describe what we have done. First, we have explored whether using edge-on discs we find a similar evolution of the size-mass relation with redshift. While using edge-on disc has some disadvantages as the effect of the internal dust is larger, it has also some important advantages as the location of the edge of the galaxy is easier to spot due to the larger contrast (higher brightness) of the disc due to the line-of-sight integration.  We show in Appendix \ref{app:edge-on} that the sizes of the edge-on discs are compatible with the sizes derived for  disc galaxies with lower inclinations reinforcing our findings. 

In Appendix \ref{app:ar}, we explore what is the effect of having galaxies with different types of inclinations in our sample. In particular, we cut out all the high inclined discs by removing all the galaxies with axis ratio lower than 0.3. This value is motivated by the fact that the inclination correction we have applied is based on a model with a fixed z$_0$/h ratio \citep[see Fig. 1 of ][]{Trujillo20}. Only for axis ratio lower than 0.3, the assumed z$_0$/h ratio can play a significant role on the inclination correction.  Our relationships do not get affected by these removal of highly inclined galaxies. This is reported on Table \ref{tab:mw-like}. There we show the R$_{\rm edge}$ and $\Sigma_{\rm edge}$  values for a MW-like galaxy (i.e. M$_{\rm stellar}$ $\sim$ 5$\times$10$^{10}$ M$_{\odot}$) according to the fits to these new filtered relationships. Those values (with the subscript \textit{ar}) are all within the 1$\sigma$ uncertainty with respect to the total sample.

Another test that we have conducted is to explore the effect of a potential contamination of spheroid-like galaxies in our sample (i.e. objects with S\'ersic index $n$ $>$ 2.5) than the visual selection of our sample can be misidentified as discs. The removal of these objects do not affect either our main results (see Appendix \ref{app:sersic} and Table \ref{tab:mw-like} parameters with the subscript \textit{n}).

Finally, and connected with our first test, we have probed whether there is any correlation between image depth and galaxy size in Appendix \ref{app:depth}. To conduct this test, we have taken advantage of the fact that the galaxies in our sample come from five different cosmological fields which are  observed with different exposure times (i.e. depth). In Fig. \ref{fig:trunc_mass_rel_depth}, galaxies are colour coded according to the depth of the images. Redder colours indicate galaxies from the deeper fields (GOODS South and North) while bluer colours correspond to the shallower fields (COSMOS, UDS and EGS). In the common stellar mass interval  (3 to 7$\times$10$^{10}$ M$_{\odot}$) for all redshift bins, we have probed whether the null hypothesis (i.e. that the samples are identical) can be rejected. By running a Kolmogorov-Smirnov test, we do not find any significant statistical evidence for different distributions among the deep and shallow samples. All the above tests give us confidence to conclude that the evolution for the R$_{\rm edge}$-mass and $\rm \Sigma_{\rm edge}$-mass relationships are not affected by any obvious systematic. 
  
For the sake of comparison,  we also included in Appendix \ref{app:effective_radii} the effective radius-mass relation for the galaxies of our sample. This relation is the one that has been used in the last years as the standard for exploring the size evolution of the galaxies. We have coloured the galaxies according to the size as provided by the radial position of the edge (i.e. R$_{\rm edge}$). As we have done with  R$_{\rm edge}$ as a proxy for galaxy size, we have quantified the evolution of r$_e$ by fixing the slope to the local value and fitting the observed relations with a power law (green lines). As previously found in the literature, the size evolution of the disc galaxies, using as proxy for the size the effective radius, is significantly smaller than using R$_{\rm edge}$. Note also how the scatter significantly grows at large redshift: from 0.15 dex in the local sample to 0.27 dex in the highest redshift bin. There is a weak tendency of galaxies with larger effective radius to be also larger using our size definition.

\begin{table*}
\centering
\caption{Best fitting values using least squares fits of the parameters of the size-mass and mass-$\Sigma_{\rm edge}$ relations   using the linear form (in the log-log space) $\rm a+bx$ . The subscript R stands for R$_{\rm edge}$ and $\rm \Sigma$ for $\Sigma_{\rm edge}$ respectively. We also add observed scatters of the relationships.}
\begin{tabular}{ccccccc}
\hline \hline
z & b$_{\rm R}$ & a$_{\rm R,b}$ & b$_{\Sigma}$ & a$_{\rm \Sigma,b}$ & $\sigma$$_{\rm R}$ & $\sigma$$_{\rm \Sigma}$ \\
\hline
Chamba et al. ($z\sim0.1$) & 0.34 $\pm$ 0.02 & -2.26 $\pm$ 0.01 & 0.32 $\pm$ 0.06  & -3.36 $\pm$ 0.02 & 0.10 $\pm$ 0.01 & 0.27 $\pm$ 0.02 \\
0.1$< \rm z <$0.3 & 0.34 & -2.36 $\pm$ 0.02 & 0.32 & -2.98 $\pm$ 0.06 & 0.11 $\pm$ 0.01 & 0.36 $\pm$ 0.05 \\
0.3$< \rm z <$0.5 & 0.34 & -2.41 $\pm$ 0.01 & 0.32 & -2.79 $\pm$ 0.04 & 0.10 $\pm$ 0.01 & 0.33 $\pm$ 0.03 \\
0.5$< \rm z <$0.7 & 0.34 & -2.47 $\pm$ 0.01 & 0.32 & -2.51 $\pm$ 0.03 & 0.11 $\pm$ 0.01 & 0.35 $\pm$ 0.02 \\
0.7$< \rm z <$0.9 & 0.34 & -2.50 $\pm$ 0.01 & 0.32 & -2.37 $\pm$ 0.02 & 0.11 $\pm$ 0.01 & 0.29 $\pm$ 0.02 \\
0.9$< \rm z <$1.1 & 0.34 & -2.55 $\pm$ 0.01 & 0.32 & -2.26 $\pm$ 0.04 & 0.13 $\pm$ 0.01 & 0.38 $\pm$ 0.03 \\
\hline
\end{tabular}
\label{tab:trunc_fits}
\end{table*}

\begin{table*}
\centering
\caption{Expected contributions to the observed scatters for the R$_{\rm edge}$-mass and $\Sigma_{\rm edge}$-mass relations due to different observational errors. We also include the expected intrinsic scatters of the observed relations ($\sigma$$_{\rm R,int}$ and $\sigma$ $_{\rm \Sigma,int}$) once they are corrected of the observational uncertainties.}
\begin{tabular}{ccccccc}
\hline \hline
z & $\sigma$$_{\rm R,mass}$ & $\sigma$$_{\Sigma,mass}$ & $\sigma$$_{\rm R,size}$ & $\sigma$$_{\Sigma,size}$ & $\sigma$$_{\rm R,int}$ & $\sigma$ $_{\rm \Sigma,int}$ \\
\hline
Chamba et al. ($z\sim0.1$) & 0.05 $\pm$ 0.01 & - & 0.07 $\pm$ 0.01 & - & 0.04 $\pm$ 0.01 & 0.27 $\pm$ 0.02 \\
0.1$< \rm z <$0.3 & 0.04 $\pm$ 0.01 & 0.09 $\pm$ 0.01 & 0.03 $\pm$ 0.01 & 0.19 $\pm$ 0.02 & 0.10 $\pm$ 0.01 & 0.29 $\pm$ 0.04 \\
0.3$< \rm z <$0.5 & 0.04 $\pm$ 0.01 & 0.09 $\pm$ 0.01 & 0.03 $\pm$ 0.01 & 0.16 $\pm$ 0.01 & 0.09 $\pm$ 0.01 & 0.28 $\pm$ 0.02 \\
0.5$< \rm z <$0.7 & 0.04 $\pm$ 0.01 & 0.09 $\pm$ 0.01 & 0.03 $\pm$ 0.01 & 0.15 $\pm$ 0.01 & 0.10 $\pm$ 0.01 & 0.31 $\pm$ 0.02 \\
0.7$< \rm z <$0.9 & 0.04 $\pm$ 0.01 & 0.09 $\pm$ 0.01 & 0.03 $\pm$ 0.01 & 0.15 $\pm$ 0.01 & 0.09 $\pm$ 0.01 & 0.24 $\pm$ 0.01 \\
0.9$< \rm z <$1.1 & 0.04 $\pm$ 0.01 & 0.09 $\pm$ 0.01 & 0.04 $\pm$ 0.01 & 0.19 $\pm$ 0.01 & 0.11 $\pm$ 0.01 & 0.31 $\pm$ 0.02 \\
\hline
\end{tabular}
\label{tab:trunc_fits_intrinsic}
\end{table*}

\begin{table*}
\centering
\caption{Values of R$_{\rm edge}$ and $\rm \Sigma_{\rm edge}$ parameters derived from our fits to galaxies with the stellar mass of the MW (i.e. $\sim$ 5$\times$10$^{10}$ M$_{\odot}$). Columns: (1) Redshift bin, (2) Number of galaxies in each bin, (3) R$_{\rm edge}$, (4) $\rm \Sigma_{\rm edge}$, (5)-(7) same as before but for the axis ratio filtered sample (i.e. excluding objects with axis ratio $<$0.3), (8)-(10) same as before but for the S\'ersic index filtered sample (i.e. excluding objects with n$>$2.5).} 
\begin{tabular}{cccccccccc}
\hline \hline
z & \# & R$_{\rm edge,MW}$ & $\rm \Sigma_{\rm edge,MW}$ & $\#_{\rm ar}$ & R$_{\rm edge,MW,ar}$ & $\rm \Sigma_{\rm edge,ar}$ & $\#_{\rm n}$ & R$_{\rm edge,MW,n}$ & $\rm \Sigma_{\rm edge,n}$ \\
 &  & (kpc) & [M$_{\odot}$ pc$^{-2}$]  & & [kpc] & [M$_{\odot}$ pc$^{-2}$] &  & [kpc] & [M$_{\odot}$ pc$^{-2}$] \\
\hline
Chamba et al. (z $\sim$ 0.1) & 254 & 23.85 $\pm$ 0.53 & 1.03 $\pm$ 0.06 & 239 & 23.83 $\pm$ 0.56 & 1.04 $\pm$ 0.06 & 254 & 23.85 $\pm$ 0.54 & 1.03 $\pm$ 0.06 \\
0.1$< \rm z <$0.3 & 64 & 19.05 $\pm$ 0.90 & 2.49 $\pm$ 0.40 & 56 & 18.68 $\pm$ 0.88 & 2.73 $\pm$ 0.45 & 41 & 19.42 $\pm$ 1.19 & 2.41 $\pm$ 0.49 \\
0.3$< \rm z <$0.5 & 138 & 17.01 $\pm$ 0.49 & 3.88 $\pm$ 0.40 & 113 & 16.42 $\pm$ 0.49 & 4.47 $\pm$ 0.45 & 77 & 17.78 $\pm$ 0.70 & 3.28 $\pm$ 0.42 \\
0.5$< \rm z <$0.7 & 282 & 14.52 $\pm$ 0.31 & 7.39 $\pm$ 0.52 & 241 & 14.23 $\pm$ 0.32 & 8.30 $\pm$ 0.57 & 175 & 15.23 $\pm$ 0.42 & 6.17 $\pm$ 0.64 \\
0.7$< \rm z <$0.9 & 343 & 13.63 $\pm$ 0.26 & 10.00 $\pm$ 0.53 & 304 & 13.45 $\pm$ 0.27 & 10.78 $\pm$ 0.61 & 228 & 14.11 $\pm$ 0.36 & 9.46 $\pm$ 0.64 \\
0.9$< \rm z <$1.1 & 221 & 12.26 $\pm$ 0.34 & 12.95 $\pm$ 1.10 & 196 & 12.13 $\pm$ 0.36 & 14.41 $\pm$ 1.22 & 152 & 12.82 $\pm$ 0.45 & 11.98 $\pm$ 1.30 \\
\hline
\end{tabular}
\label{tab:mw-like}
\end{table*}

\subsection{The cosmic size evolution of disc galaxies}
\label{subsec:evolution}

\begin{figure*}
\centering
    \includegraphics[width=\textwidth]{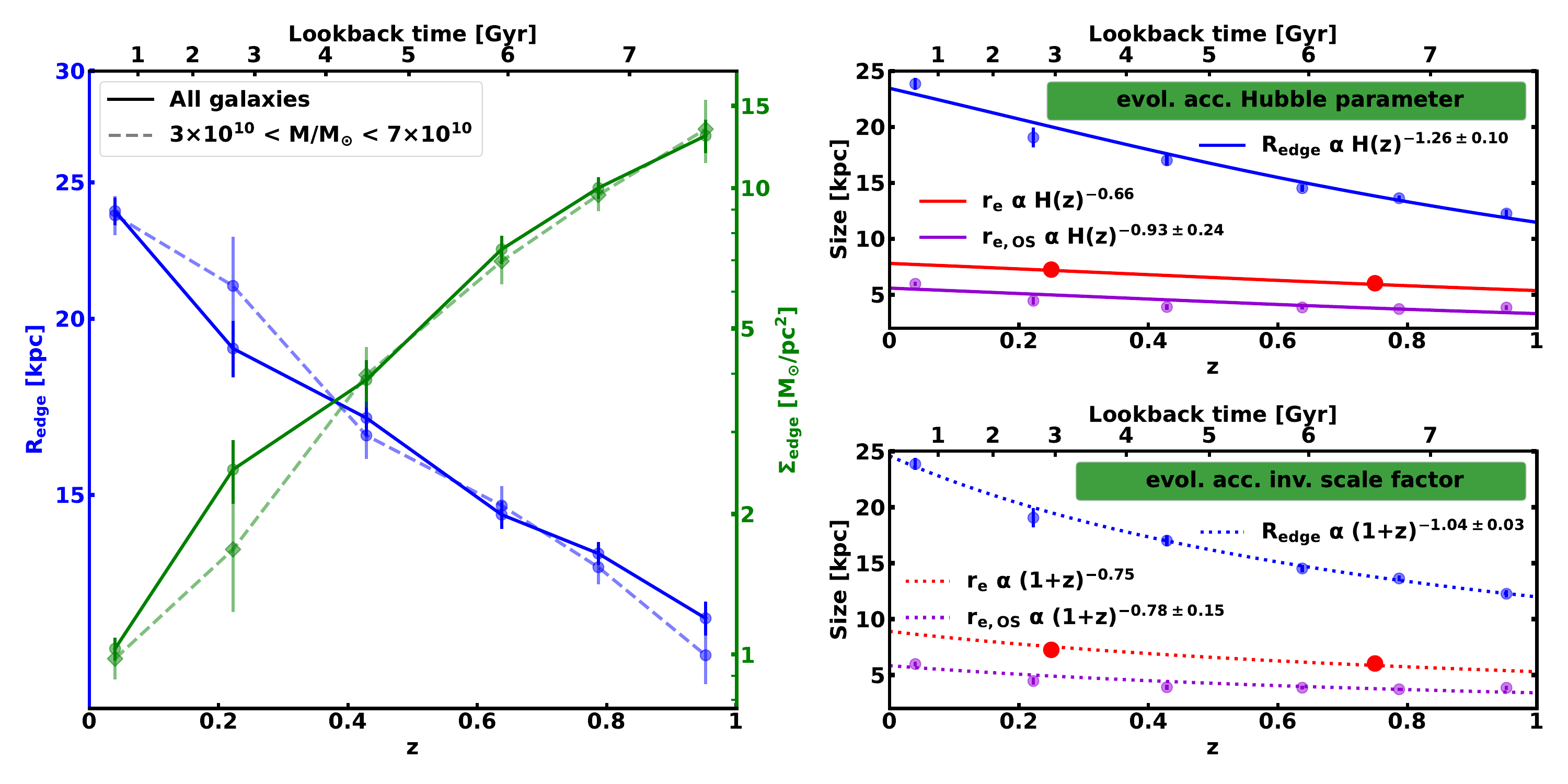}
    \caption{Left column: Size evolution of disc galaxies using as a proxy the radial location of the edge of the star forming discs (R$_{\rm edge}$; see left blue axis). We also show the evolution of the  stellar  mass surface density at the edge  ($\rm \Sigma_{\rm edge}$; see right green axis). The solid  lines show the evolution corresponding to a disc galaxy with the stellar mass of the MW (i.e. M$_{\rm stellar}$ $\sim$ 5$\times$10$^{10}$ M$_{\odot}$) according to the best linear fitting to the entire population (see text for details). The dashed lines correspond to the average properties of galaxies within the stellar mass range 3$\times$10$^{10}$ < M/M$_{\odot}$ < 7$\times$10$^{10}$. The redshift of each data point corresponds to the median value of the galaxy redshifts for each redshift bin. MW-like galaxies increase on average their sizes by a factor of $\sim$2 since z=1. The  stellar mass surface density at the end of their star forming discs ($\rm \Sigma_{\rm edge}$) decreased by an order of magnitude since that epoch. Right column: Evolution of the size of the galaxies in our sample using as a size indicator either the edge of the star forming discs (R$_{\rm edge}$; blue points) or its effective radius (r$_{\rm e,OS}$; purple points with the subscript OS indicating the galaxies of Our Sample). In the upper panel this evolution is fitted using a power-law function of the Hubble parameter H(z) while in the lower panel we use a power-law function of the inverse of the scale factor (i.e. (1+z)). We add for comparison the evolution of the effective radius for the general population of star forming galaxies given in \citet{vanderWel14}.}
    \label{fig:evol_average_gals}
\end{figure*}

In order to illustrate the size evolution of the disc galaxies since z$\sim$1, we have quantified the size (R$_{\rm edge}$) evolution at a fixed stellar mass using two independent ways. In the first case, we use the global fitting to the size-mass relations as a reference. We derive the size of a galaxy with a stellar mass of the MW (i.e. M$_{\rm stellar}$ $\sim$ 5$\times$10$^{10}$ M$_{\odot}$) according to the best fitting result. The cosmic size evolution of a disc galaxy with such stellar mass is given in the left panel of Fig \ref{fig:evol_average_gals}  (solid blue line).  The other way to illustrate the cosmic size evolution is by calculating the average size of all the galaxies in a common stellar mass bin (3$\times$10$^{10}$ < M/M$_{\odot}$ < 7$\times$10$^{10}$) for the different redshift bins. Note that we do not use the entire mass range  as the different redshift bins (which covers a different cosmic volume) have galaxies with a different mass distribution. To avoid a bias towards low mass (at low redshift) or high mass (at high redshift) we stick to the previous (and common) stellar mass range at all redshifts. The size evolution of the galaxies in that mass bin is shown in the left panel of Fig \ref{fig:evol_average_gals}  (dashed blue line). We repeat the same exercise for the value of the stellar surface mass density ($\rm \Sigma_{\rm edge}$) at the radial position of the edge. The evolution of the above two quantities is similar (within the error bars) using the two different approaches.

The typical size evolution of a disc galaxy using R$_{\rm edge}$ as a size proxy is a factor of $\sim$2 since z=1 (i.e. from 12 kpc some 8 Gyr ago to 24 kpc for a present-day MW-like galaxy). Conversely, the stellar surface mass density at the edge position increases an order of magnitude with redshift. To be more precise, $\rm \Sigma_{\rm edge}$ increases from a value of $\sim$1 M$_{\odot}$ pc$^{-2}$ at z=0  to $\sim$13 M$_{\odot}$ pc$^{-2}$ at z=1. Disc galaxies with the stellar mass of the MW were fundamentally different 8 Gyr ago than nowadays: the extension of  their star forming discs were a factor of 2 smaller and their stellar mass surface densities at the end of the star forming disc were a factor of 10 higher.

In the right panels of Fig. \ref{fig:evol_average_gals}, we quantify the galaxy size evolution using both R$_{\rm edge}$ and r$_e$. We parameterized this evolution as a function of the Hubble parameter H(z) (top plot, solid lines) and the inverse of the cosmic scale factor, i.e. (1+z) (bottom plot, dotted lines). The  evolution of the galaxy size is calculated using the results of the fitting to the entire population (i.e. the values shown in the solid blue line in the left plot). The red data points correspond to the evolution of the effective radius published by \citet{vanderWel14} (their Table 1 and Figure 6) for their star-forming sample. The purple points show the  evolution of the effective radius but this time only using the galaxies analysed in this work (OS = Our Sample). 

The relative evolution of the effective radius of the galaxy sample selected by their star formation activity \citep[i.e.][]{vanderWel14} --red lines-- and the effective radii for our morphologically selected sample --purple lines-- is similar. However, the effective radius of the former sample is around a factor of 2 larger in r$_e$ than the visually classified as discs. There might be a number of reasons for this discrepancy. First, the stellar mass range and the photometric masses assumed in both samples are not the same. Second, the colour-selected (i.e.star-forming) galaxies probably have prominent discs and therefore, the effective radius is larger than for objects that could host more passively evolving populations. 

 The main message of  Fig. \ref{fig:evol_average_gals} is that the size evolution of the disc galaxies using R$_{\rm edge}$ as the proxy for galaxy sizes is stronger than the observed evolution of their effective radius. Another important outcome of the  size evolution we have found is that it can be very well represented as a simple function of the evolution of the  scale factor (i.e. (1+z)$^{-1}$). In fact, we find a best fitting function of (1+z)$^{-1.04\pm0.03}$. This increase by a factor of 2 in the size of the MW-like galaxies since z=1 is more in line with the expected size evolution of disc galaxies based on simple theoretical predictions \citep[see e.g.][]{Mo98}.

\subsection{The disc growth of MW-like galaxies}
\label{subsec:velocity}

We can provide a rough estimate of the speed of the radial evolution of the star forming radius (i.e. the in-situ growth of the discs) with cosmic time under the assumption that disc galaxies do not evolve significantly in stellar mass since z = 1. Based on the difference between R$_{\rm edge}$ nowadays and at z=1, the average speed growth we derive for a disc galaxy with the stellar mass of the MW  ($\sim$ 5$\times$10$^{10}$ M$_{\odot}$) is 1.60$\pm$0.45 kpc Gyr$^{-1}$. For individual galaxies, which certainly are evolving in stellar mass since z = 1, the above value is a lower limit to the average speed growth of the star forming disc.

Interestingly, the above number can be compared with the estimation of the growth speed for two present-day MW-like galaxies (NGC4565 and NGC5907). \citet{Martinez-Lombilla19} found a present-day growth for these two galaxies of 0.6-1 kpc Gyr$^{-1}$. These values are within a factor 2 to 3 the crude estimation we have done in the present work. 

Finally, it is worth comparing the size evolution we found in this work with previous attempts of measuring this quantity using break signatures on the surface brightness profiles. \citet{Trujillo05} and \citet{Azzollini08} did that exercise using the break radius. It should be noted that at the time these works were carried out, the break radius and the truncation of the profiles were interchangeably called the same. Currently, we know the break radius reflects a signature well inside the star-forming discs (normally associated to the end of the main spiral arms) while the edge of the star forming discs corresponds to a decline on the number of stars outside the star forming radius \citep[see e.g.][]{Martin-Navarro12}.  According to \citet{Azzollini08} the location of the break radius evolved by 0.4 kpc Gyr$^{-1}$. Most of the galaxies in \citet{Azzollini08} sample are less massive than the ones studied here. Therefore, it is not straightforward to conclude whether the evolution of break and the edge radius is similar or different with cosmic time. The analysis of that issue is beyond the scope of this paper.


\section{Conclusions}
\label{sec:conclusions}

We present the size evolution since z = 1 for MW-like galaxies using a physically motivated size proxy based on the radial extension of the in-situ star formation \citep{Trujillo20}. Following the methodology prescribed by \citet{Chamba22}, to estimate such extension, we look for truncations on the stellar surface mass density and surface brightness profiles of the galaxies which are indicative of a sudden drop on the star formation activity at that radial distance. To carry out this work, we have used the deepest available HST cosmic field observations. We have explored a total of 1048 massive (M$_{\rm stellar}$ $>$ 10$^{10}$ M$_{\odot}$) disc galaxies with high-quality spectroscopic redshifts.

To characterise the galaxy sizes (R$_{\rm edge}$) and the stellar mass surface density at the position where the edge is located ($\Sigma_{\rm edge}$), we have conducted a careful detailed analysis of the galaxy surface brightness profiles (both azimuthally averaged and along the semi-major axis). In addition to the surface brightness profiles in all the bands available for these galaxies, we have extracted colour and stellar mass density profiles. The quantities we provide are corrected by Galactic dust extinction, inclination and cosmological dimming.

By using R$_{\rm edge}$, we find that disc galaxies with stellar masses similar to MW (i.e. $\sim$ 5$\times$10$^{10}$ M$_{\odot}$) have increased their size by a factor of 2 since z=1. The  stellar mass surface density at the position of the edge, on the opposite, has decreased by a factor of 10 since that epoch. This decrement by an order of magnitude is potentially linked to the same decrease in the star formation rate since that time \citep[][and references therein]{Madau14}. The lowering in the star formation activity with cosmic time would decrease the released energy on the available gas, therefore allowing lower gas densities to collapse and form stars progressively farther away in the outer parts of the discs. The strong evolution of the stellar mass surface density at the edge of the disc galaxies since z=1 contrasts sharply with the evolution of the central regions, which remain essentially unchanged since that time (see Appendix \ref{app:central_mass_profiles}).

The picture that emerges of the size evolution of disc galaxies by using  R$_{\rm edge}$  is significantly different from the one using the effective radius.  For instance, there have been recent claims about a negligible size (i.e. effective radius) evolution for MW-like galaxies in the last 10 Gyr \citep{Hasheminia22}. These results are not in contradiction with our findings. This is because the effective radius is driven by the galaxy light concentration (i.e. the presence or not of a prominent bulge) and does not reflect the intuitive extension of the galaxy (cf. its disc) as  R$_{\rm edge}$ does. 

There are, of course, a number of caveats in the present study. First, we have not corrected the effect of internal dust obscuration in the surface brightness profiles of our galaxy sample. This could affect our determination of the total stellar mass and the  stellar mass surface density at the edge, especially for highly inclined galaxies. We stress, however, that at least for the local MW-like galaxies seen edge-on such as  NGC4565 and NGC5907, the amount of dust at the edge position is not significant \citep[compared to the internal regions, see e.g.][]{Martinez-Lombilla19}.  Second, our sample is composed of (fairly massive) galaxy discs that, by the way they are selected, avoiding interactions and contamination by luminous neighbouring objects, might be biased towards low-density environments. Finally, even though we make use of HST observations, the effect of the PSF complicates the accuracy at estimating the size and axis ratio  (0.2\arcsec\ PSF FWHM in the reddest bands translates into 1.4 kpc spatial resolution at $z$ = 1) for the smaller and higher redshift galaxies in our sample.

We would like also to remark the finding of a large number of massive (M$_{\rm stellar}$ $>$ 5$\times$10$^{10}$ M$_{\odot}$) and compact (R$_{\rm edge}$ $<$10 kpc) discs found at z=1 that are not found in the local Universe (see Fig. \ref{fig:trunc_mass_rel_colored_sigma}). These very dense mini-disc galaxies  gradually disappear from our sample as the cosmic time progresses. The disappearance of these objects at low redshift remind us that, in order to characterise the size evolution of individual objects, we need to understand in detail the so-called progenitor bias \citep[new galaxies entering the selection criteria at later epochs, i.e.][]{Carollo13,Shankar15,Zanisi21}.

Looking ahead to the immediate future, the very deep, highly spatially resolved and near-infrared images provided by the James Webb Space Telescope (JWST) will be instrumental for the purpose of better characterising the size of the galaxies (see some examples in Appendix \ref{app:jwst}) even at redshifts beyond z=1.  Additionally, next generation synoptic facilities, such as the Euclid, Roman and Rubin telescopes, will also deliver  the required depths $\sim$29-30 mag/arcsec$^{2}$ \citep[3$\sigma$ in 10$\times$10 arcsec boxes, see e.g.][]{Borlaff22,Scaramella22,Martin22} to conduct these studies with much larger statistics \citep[see a preview of the expected results at such a depth in][]{2021A&A...654A..40T}.

\begin{acknowledgements}

We are grateful to the anonymous referee, who carried out a very careful review of the manuscript and provided a number of suggestions that improved the clarity and robustness of the present work. We thank Nushkia Chamba for very valuable input to build this work and for sharing in advance the results of the local sample we have used as a reference sample. We are indebted to Mohammad Akhlaghi and Ra\'ul Infante-S\'aiz for their Gnuastro \citep{Akhlaghi15,Akhlaghi19} routines, especially the code rgb-asinh, without which the galaxy colour images would not look the same. We thank Arjen van der Wel for providing in advance the redshifts for LEGA-C DR3. We also are  very grateful to Juan E. Betancort, Israel Matute, Diego S\'aez-Chill\'on, Peter Erwin, Ignacio Ferreras, Alex Vazdekis, Anna Ferr\'e-Mateu, Jes\'us Falc\'on-Barroso, John Peacock, Samane Raji, Javier Mart\'in-Campo and Javier Blasco for their help in several aspects of this paper.

F.B. acknowledges support from the grants PID2020-116188GA-I00 and PID2019-107427GB-C32 by the Spanish Ministry of Science and Innovation. F.B. also acknowledges the support by FCT via the postdoctoral fellowship SFRH/BPD/103958/2014. This work is supported by Funda\c{c}\~ao para a Ci\^encia e a Tecnologia (FCT) through national funds (UID/FIS/04434/2013) and by FEDER through COMPETE2020 (POCI-01-0145-FEDER-007672). I.T. acknowledges support from the ACIISI, Consejer\'{i}a de Econom\'{i}a, Conocimiento y Empleo del Gobierno de Canarias and the European Regional Development Fund (ERDF) under grant with reference PROID2021010044 and from the State Research Agency (AEI-MCINN) of the Spanish Ministry of Science and Innovation under the grant PID2019-107427GB-C32 and IAC project P/300624, financed by the Ministry of Science and Innovation, through the State Budget and by the Canary Islands Department of Economy, Knowledge and Employment, through the Regional Budget of the Autonomous Community.

We have used extensively the following software packages: TOPCAT \citep{Taylor05}, ALADIN \citep{Bonnarel2000} and MATPLOTLIB \citep{matplotlib_ref}.
This research made use of APLpy, an open-source plotting package for Python \citep{aplpy_ref}.
This work has made use of the Rainbow Cosmological Surveys Database, which is operated by the Centro de Astrobiolog\'ia (CAB/INTA), partnered with the University of California Observatories at Santa Cruz (UCO/Lick,UCSC).
Based on zCOSMOS observations carried out using the Very Large Telescope at the ESO Paranal Observatory under Programme ID: LP175.A0839.

\end{acknowledgements}


\bibliographystyle{aa.bst}
\bibliography{refs.bib}


\begin{appendix}

\section{Comparison with JWST}
\label{app:jwst}

Deep JWST data from the GOODS-South region have recently been published \citep{2023arXiv230602465E,2023arXiv230602466R}. These data (of greater depth and better spatial resolution than those used in the present paper) allow us to make a quick comparison with the HST data and thus to explore the robustness of our estimate of galaxy sizes. They also give us an indication of what to expect in future work to measure the size of distant galaxies using the techniques described in this paper.

\begin{figure*}
\centering
    \includegraphics[width=0.7\textwidth]{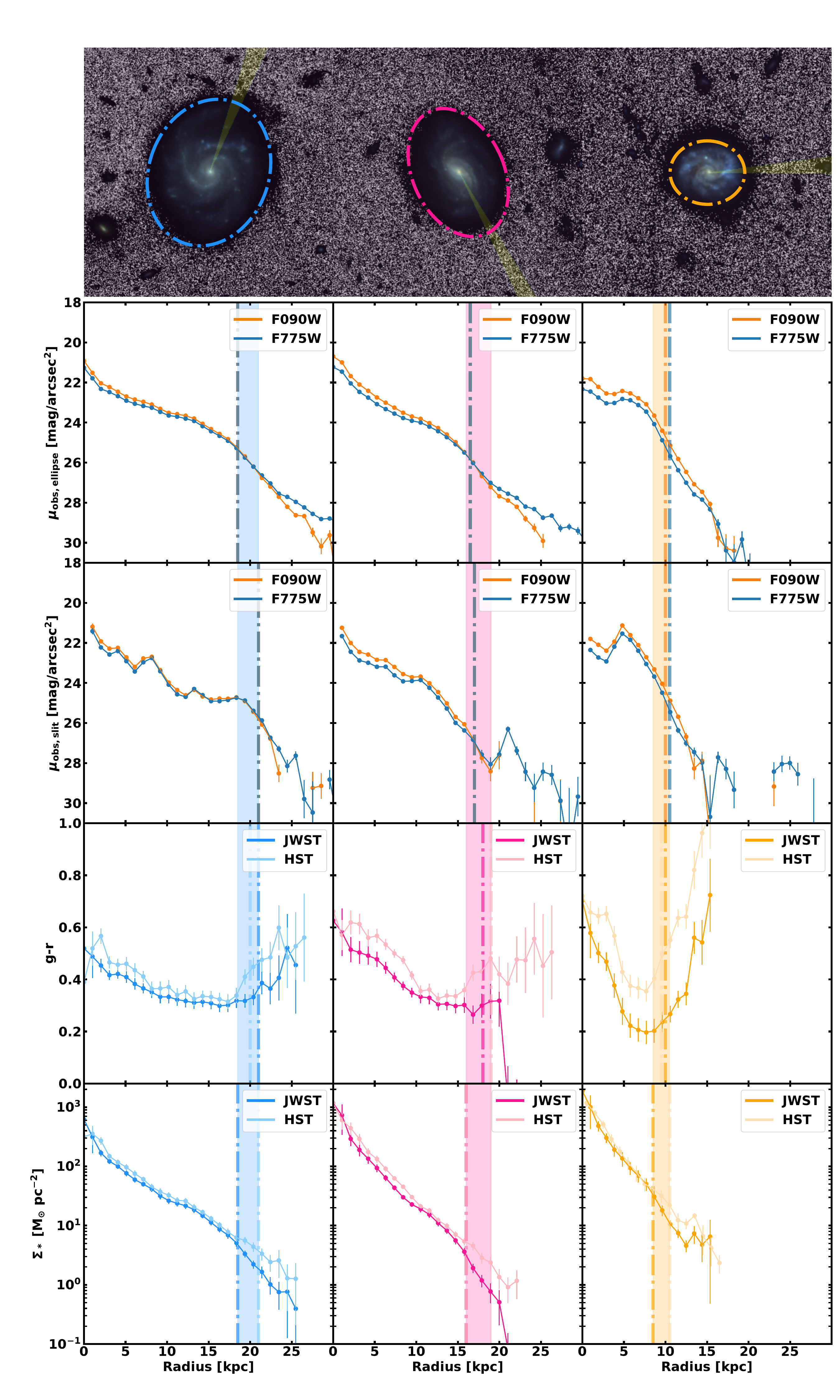}
    \caption{Comparison between HST and JWST datasets for a sample of three MW-like galaxies with similar stellar masses (M$_{\rm stellar}$$\sim$5$\times$10$^{10}$ M$_{\odot}$) at different redshifts (from left to right): A) 17058\_GOODSS at $z$ = 0.6218 in blue, B) 12634\_GOODSS at $z$ = 0.6684 in magenta, and C) 12887\_GOODSS at $z$ = 0.997 in dark yellow. The first row shows their colour RGB images (based on JWST data; 70$\times$70 kpc), with their sizes (R$_{\rm edge}$) highlighted by dotted dashed lines, whose positions are also marked in the plots below. The orientation of the ``slit'' we used to extract the profiles plotted in the panels of the third row is shown in yellow. The second row shows the surface brightness profiles in F775W (HST) and F090W (JWST) using azimuthally averaged profiles (``ellipse method''). The third row is similar to the second, but this time following the semi-major axis profiles (``slit method'').  The fourth row shows the $g-r$ colour profiles derived with each telescope, and the fifth row shows the stellar surface mass density profiles based on HST and JWST. In each panel, the vertical dashed lines indicate the position of R$_{edge}$ according to the methodology proposed in this paper. The coloured vertical regions indicate the uncertainty at measuring the position of the edge of the galaxy using independently the information given in each panel.}
    \label{fig:jwst}
\end{figure*}

In Fig. \ref{fig:jwst}, we show the detailed analysis of three galaxies located in the GOODS-South. All these galaxies have a similar stellar mass of M$_{\rm stellar}$$\sim$5$\times$10$^{10}$ M$_{\odot}$. The colour images of the galaxies are produced using the JWST images, which have an exquisite spatial resolution, allowing us to study the internal structure of the discs of these galaxies in great detail. In addition to the images, we show the surface brightness profiles using both the ellipses and the slit along the major axis. To facilitate comparison with the HST data, we show only the surface brightness profile in the F775W (HST) and F090W (JWST) bands due to their  wavelength and depth proximity in this particular field. We also show the rest-frame g-r colour using the set of bands from each telescope separately and the surface stellar mass profile.

In each panel, we show the location of R$_{\rm edge}$ using only the information contained in each panel. We also do this separately for HST and JWST. The superior data quality of JWST allows us to identify the galaxy edges more clearly. This is particularly evident in the stellar surface mass density and colour profiles. The increase in spatial resolution (and therefore narrower PSF) of JWST relative to HST at a given wavelength (in this case around the observed 8500\AA) reduces the amount of light in the outer part of the galaxy, making the surface brightness and stellar mass density profiles steeper beyond R$_{\rm edge}$. As a result, the h1/h2 ratio (i.e. the ratio between the inner and outer disc slopes) is larger in JWST than in HST, and the R$_{\rm edge}$ is easier to identify. In general, we find good agreement in the estimation of R$_{\rm edge}$ using both the different methods and facilities. The typical uncertainty in these examples is around $\pm$1 kpc.

\section{Size (R$_{\rm edge}$) and stellar surface mass densities at the edge  ($\Sigma_{\rm edge}$)  in edge-on galaxies}
\label{app:edge-on}

The brighter surface brightness of highly inclined discs due to the line-of-sight integration  facilitates the location of the disc edge (and therefore the size determination) in these systems. For this reason, a good way of testing the accuracy of our results is to probe  where visually identified edge-on systems are distributed in the size - mass and stellar surface mass density - mass relations compared to the galaxies with lower inclinations of our sample.

\begin{figure*}
\centering
    \includegraphics[width=\textwidth]{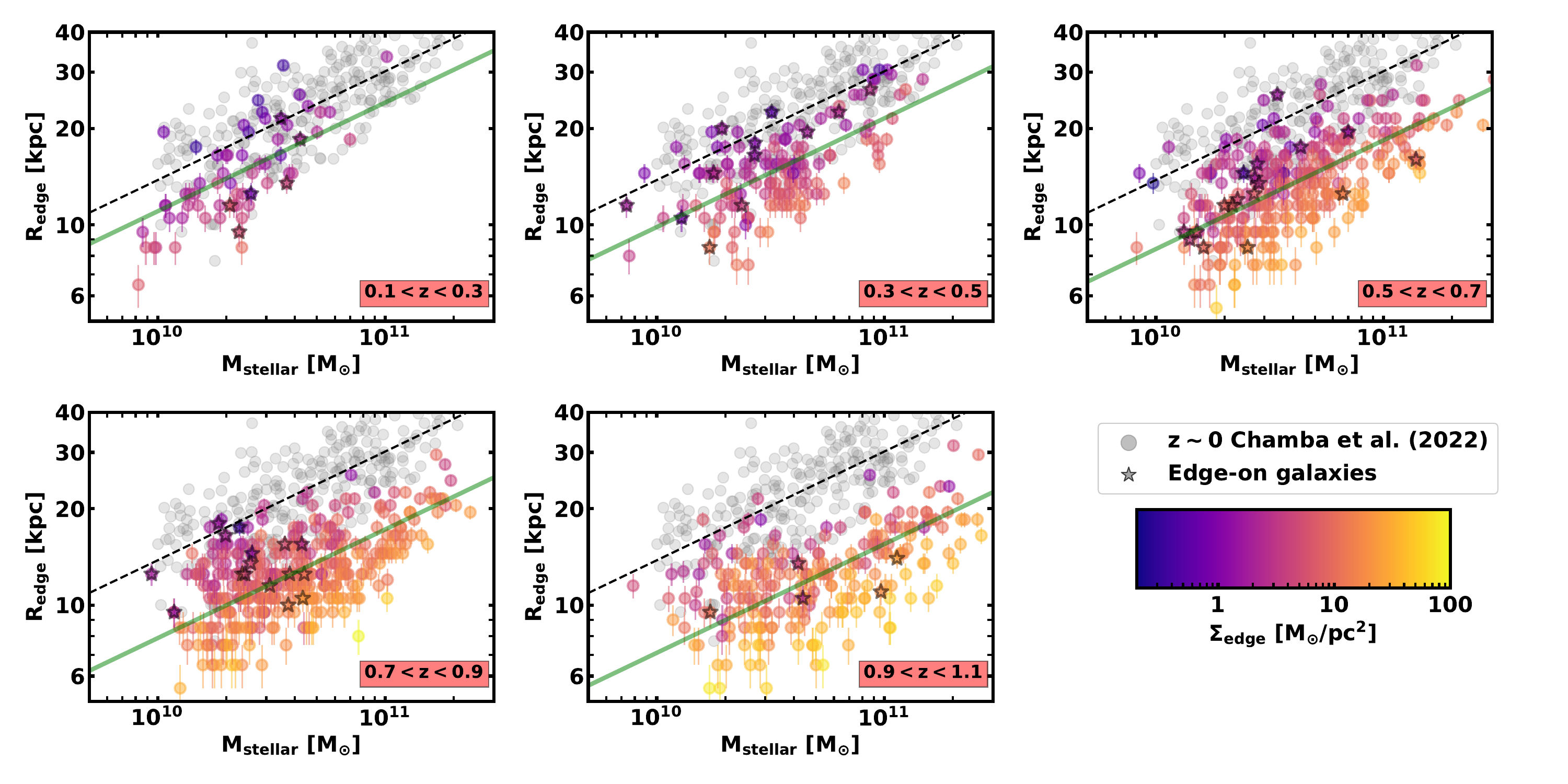} 
    \caption{Same as Fig. \ref{fig:trunc_mass_rel_colored_sigma} but this time highlighting the edge-on galaxies (denoted by stars) on top of the general sample. Both samples of galaxies share a similar size-mass distribution.}
    \label{fig:trunc_mass_rel_colored_sigma_with_edge_on_gals}
\end{figure*}

\begin{figure*}
\centering
    \includegraphics[width=\textwidth]{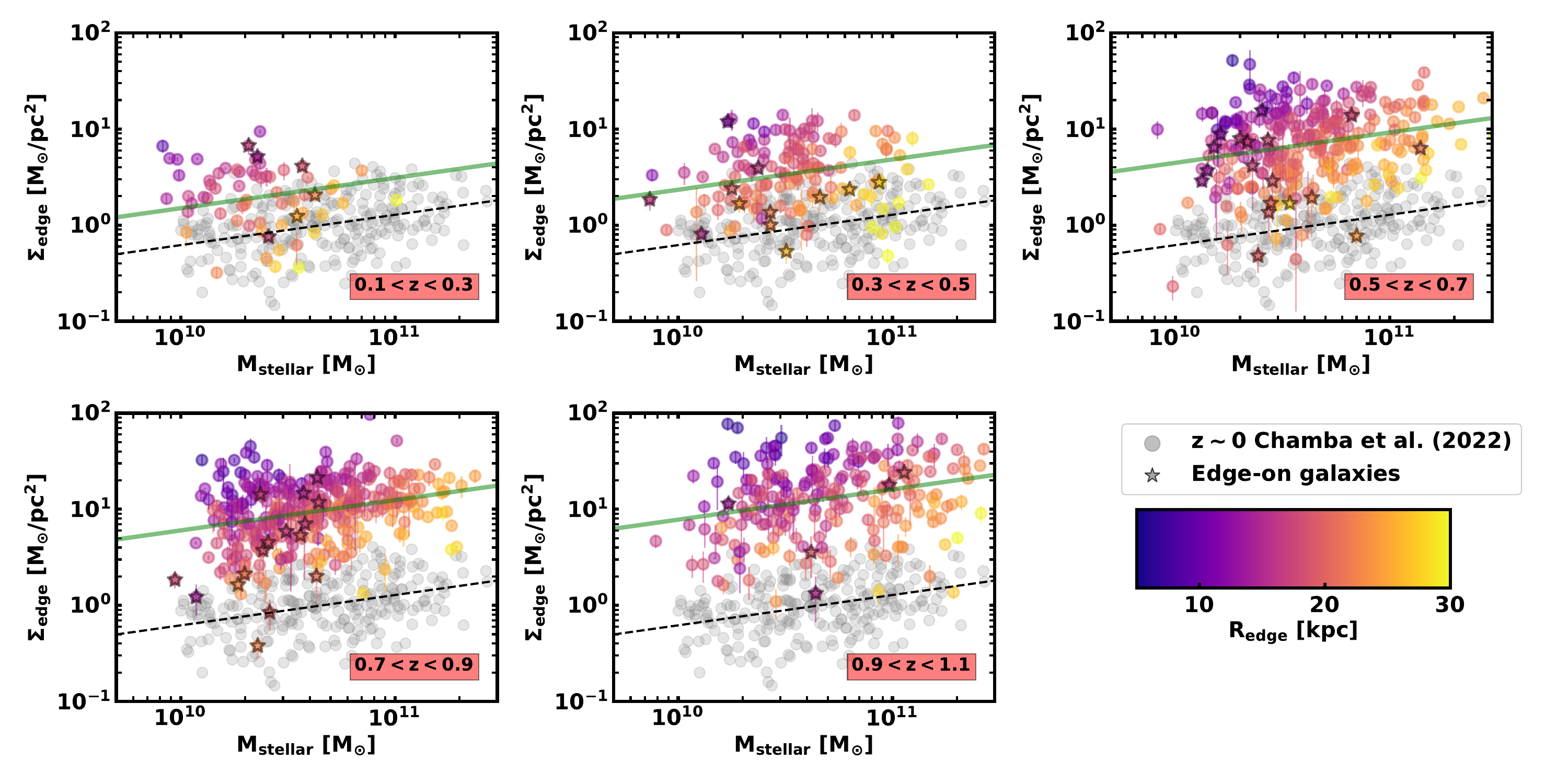} 
    \caption{Same as Fig. \ref{fig:sigma_mass_rel_colored_trunc} but this time highlighting the edge-on galaxies (denoted by stars) on top of the general sample. Both samples of galaxies share a similar stellar surface mass density - mass distribution.}
    \label{fig:sigma_mass_rel_colored_trunc_with_edge_on_gals}
\end{figure*}

We show R$_{\rm edge}$ - mass and the $\Sigma_{\rm edge}$ - mass relations in Fig.  \ref{fig:trunc_mass_rel_colored_sigma_with_edge_on_gals} and \ref{fig:sigma_mass_rel_colored_trunc_with_edge_on_gals} respectively. Edge-on galaxies are plotted using stars symbols. We find a general good agreement on the distribution of the edge-on sample and our main sample in these two plots. There is a small tendency to edge-on systems to be a bit larger (and therefore having lower values of $\Sigma_{\rm edge}$) than low inclined galaxies. However, it is worth noticing that in edge-on systems, due to the blurring effect of the Point Spread Function, the visual selection of disc galaxies is more prone to select  the larger systems.

\section{The effect of removing highly inclined galaxies from the sample}
\label{app:ar}

Highly inclined galaxies are more likely (due to the effect of internal dust and uncertainty at determining the inclination) to have both the total stellar mass and stellar surface mass density less accurately determined than low inclined systems. To explore the effect of these galaxies  in the size-mass and stellar surface mass density - mass relation, we remove the objects with axis ratio $<$ 0.3 (138 in number) from our analysis in this Appendix section.  

\begin{figure*}
\centering
    \includegraphics[width=\textwidth]{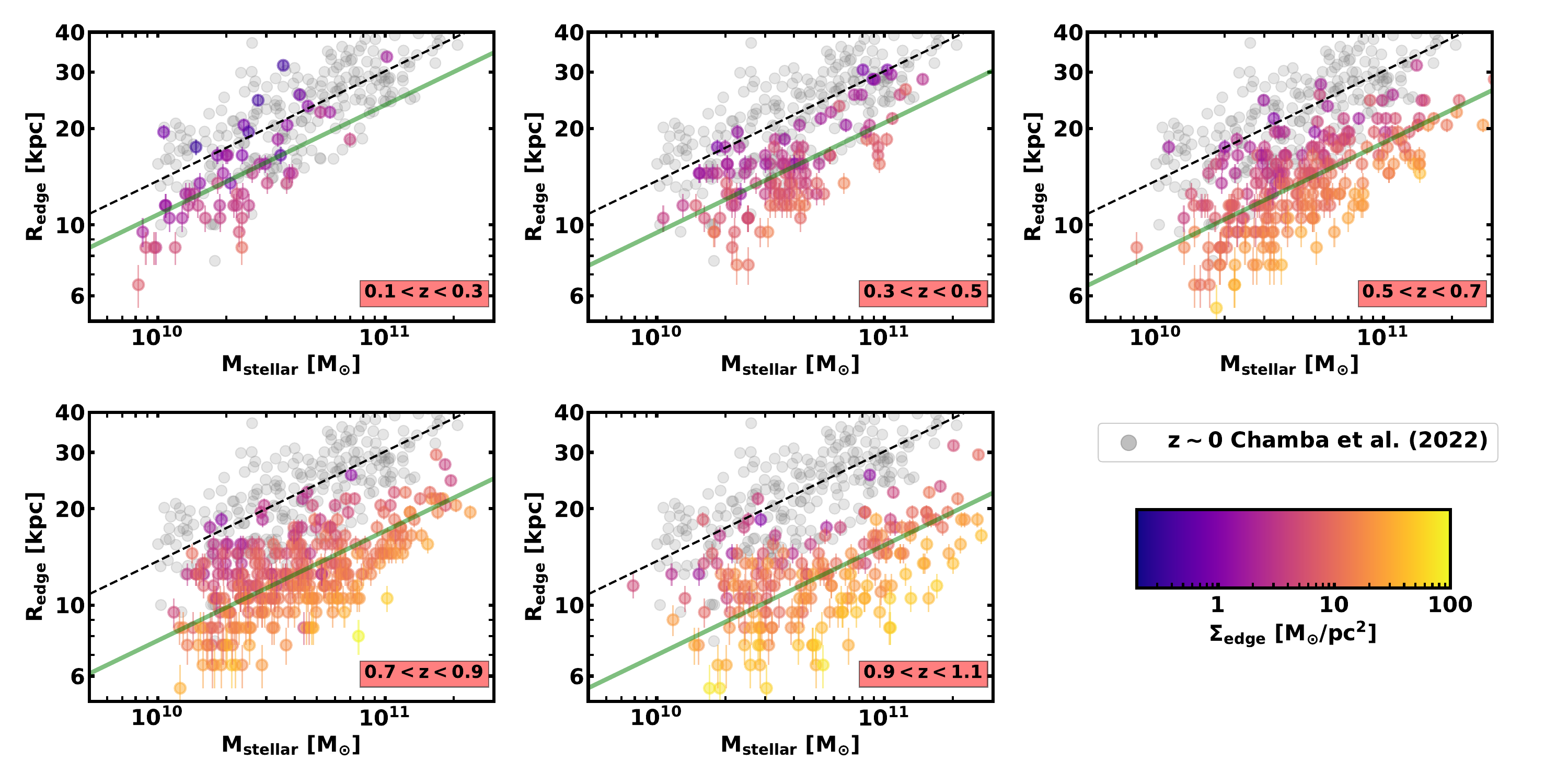}
    \caption{Same as Fig. \ref{fig:trunc_mass_rel_colored_sigma} but this time removing highly inclined galaxies with axis ratio $<$ 0.3.}
    \label{fig:trunc_mass_rel_filtered_by_axis_ratio}
\end{figure*}

\begin{figure*}
\centering
    \includegraphics[width=\textwidth]{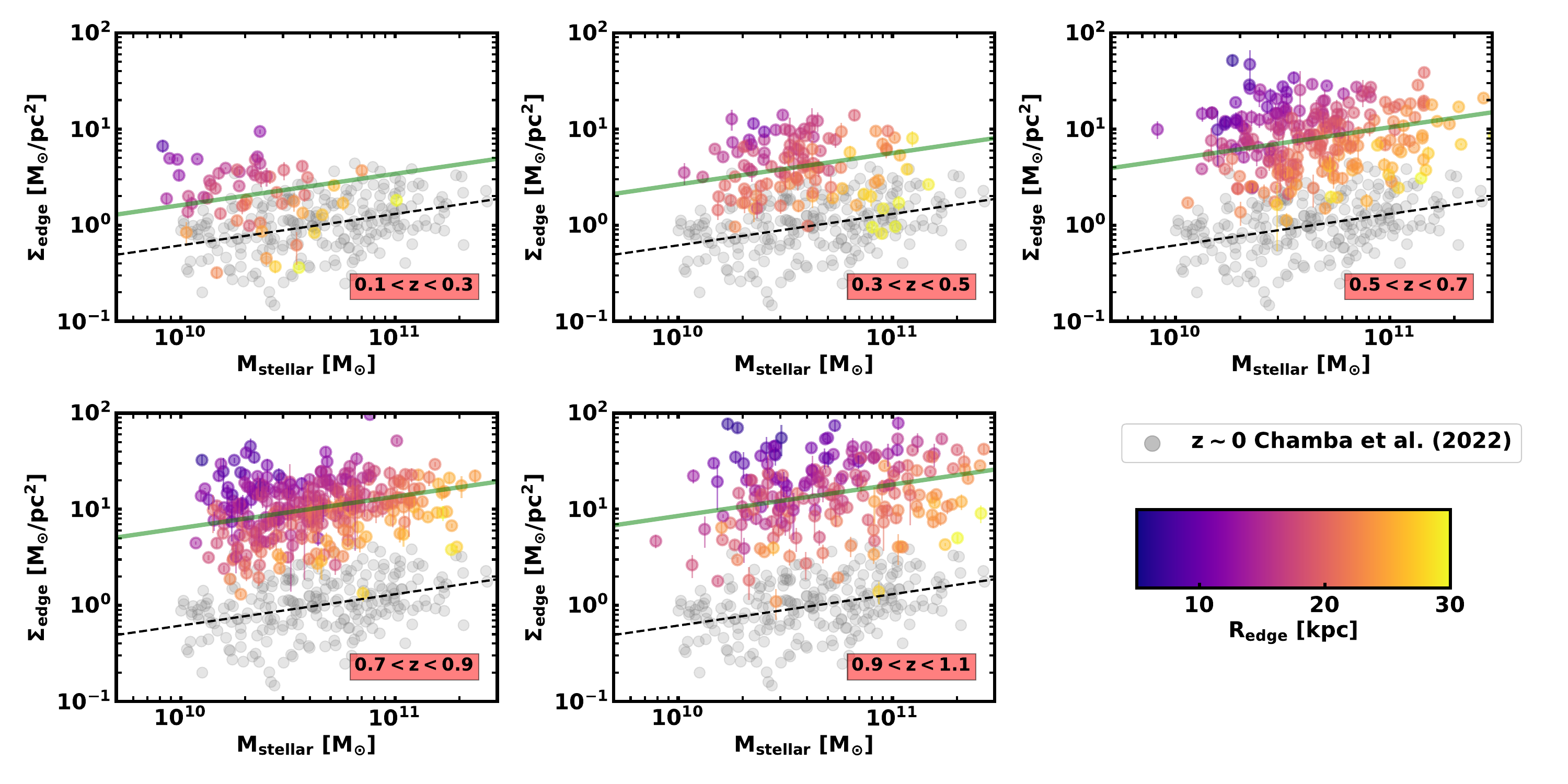}
    \caption{Same as Fig. \ref{fig:sigma_mass_rel_colored_trunc} but this time removing  highly inclined  galaxies with axis ratio $<$ 0.3.}
    \label{fig:sigma_mass_rel_filtered_by_axis_ratio}
\end{figure*}

The R$_{\rm edge}$ - mass and the $\Sigma_{\rm edge}$ - mass relations without the highly inclined galaxies can be found in \ref{fig:trunc_mass_rel_filtered_by_axis_ratio} and \ref{fig:sigma_mass_rel_filtered_by_axis_ratio}. The Table \ref{tab:mw-like} (columns 5 to 7) display the values of least-squares fit in this case for MW analogues. The new observed scatters for these relationships are also very similar (the mean for all redshift bins are 0.11 and 0.31 dex respectively) than the ones derived for the general sample. The main conclusion is that our analysis does not change by removing or not these highly inclined galaxies.

\section{Filtering our sample by the S\'ersic index}
\label{app:sersic}

The selection of our disc galaxies is purely based on the morphological appearance using automatic methods. Nonetheless, 375 galaxies with global S\'ersic indices $n$ greater than 2.5 (according to \citet{vanderWel12} measurements) are part of our sample. These galaxies would have been traditionally avoided in other works that selected late-type objects based on surface brightness parametric fitting. To test the impact on the addition of these galaxies, we have removed them in the present Appendix.

\begin{figure*}
\centering
    \includegraphics[width=\textwidth]{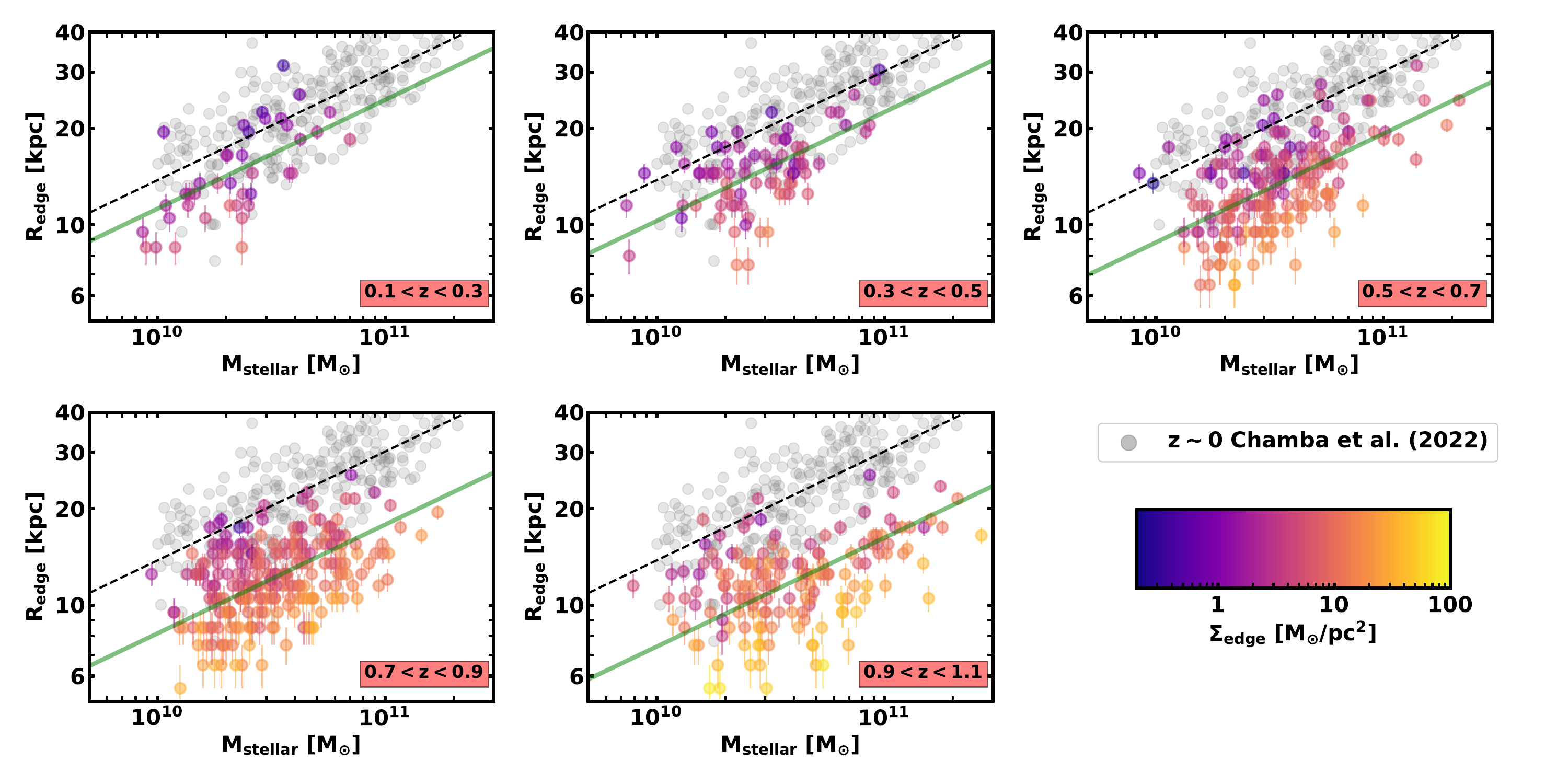}
    \caption{Same as Fig. \ref{fig:trunc_mass_rel_colored_sigma} but this time removing galaxies with S\'ersic index n $>$ 2.5.}
    \label{fig:trunc_mass_rel_filtered_by_sersic}
\end{figure*}

\begin{figure*}
\centering
    \includegraphics[width=\textwidth]{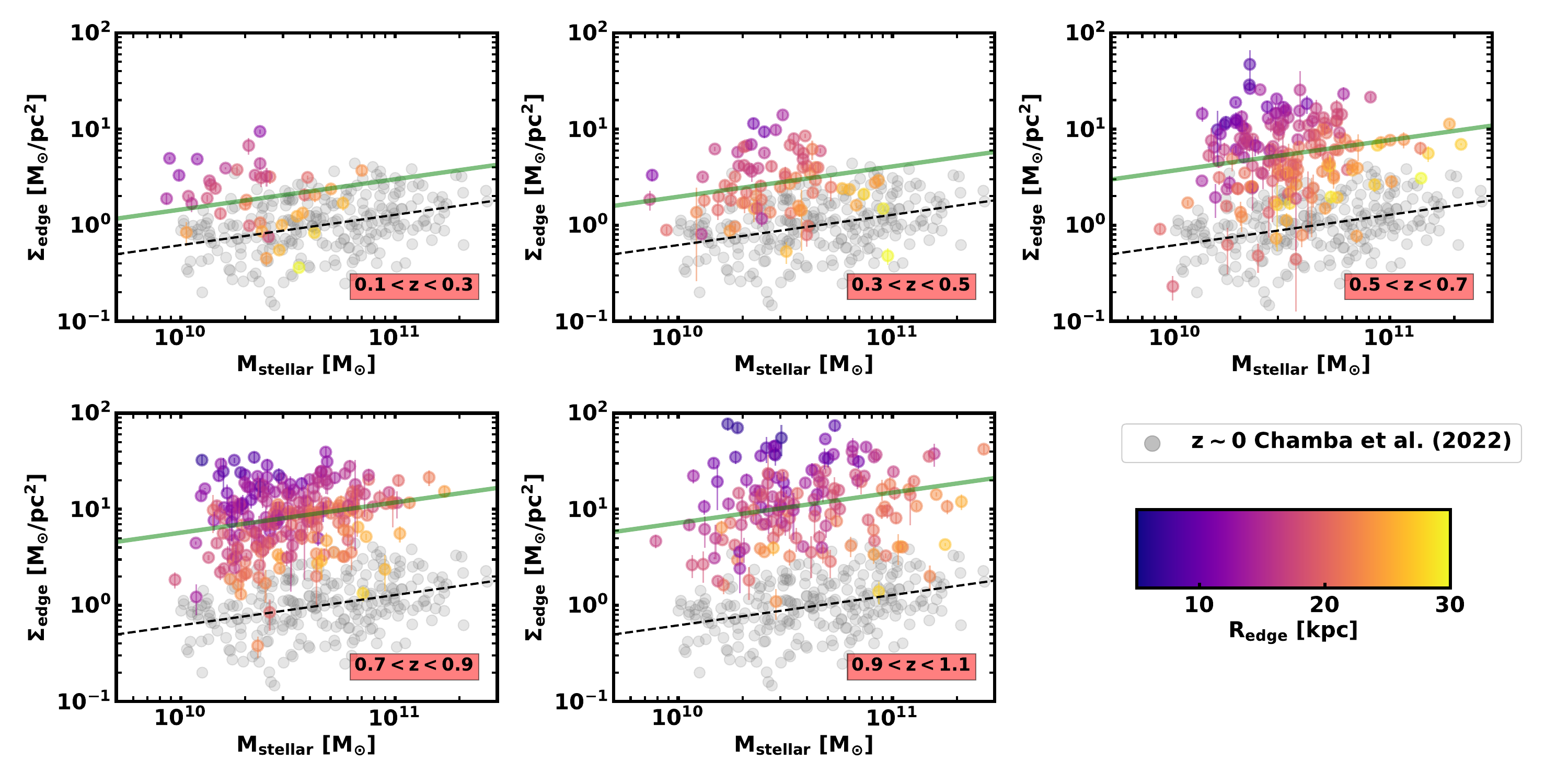}
    \caption{Same as Fig. \ref{fig:sigma_mass_rel_colored_trunc} but this time removing galaxies with S\'ersic index n $>$ 2.5.}
    \label{fig:sigma_mass_rel_filtered_by_sersic}
\end{figure*}

The R$_{\rm edge}$ - mass and the $\Sigma_{\rm edge}$ - mass relations   without these high-S\'ersic index galaxies can be found in Fig. \ref{fig:trunc_mass_rel_filtered_by_sersic} and Fig. \ref{fig:sigma_mass_rel_filtered_by_sersic}. The Table \ref{tab:mw-like} (columns 8 to 10) display the values of least-squares fit in this case for MW analogs. The new observed scatters for these relationships are also very similar (the mean for all redshift bins are 0.11 and 0.34 dex respectively) than the ones derived for the general sample. We do not appreciate any important effect resulting from the inclusion or not of objects with n$>$2.5 in our sample.

\section{Effect on characterising the size of the galaxy depending on the image depth}
\label{app:depth}

As the size of the galaxies depends on the ability of detecting the edge on the stellar mass surface density and surface brightness profiles, one may wonder whether the depth of the images in the present study plays any role at detecting the position of this low surface brightness feature. To answer this question, we have explored whether the size - mass relation changes depending on the cosmic fields used to measure the galaxy size. The fields with deeper observations are GOODS-South and GOODS-North, while the shallower ones are COSMOS, UDS and EGS. The median differences in limiting surface brightness for these two kinds of fields are on average 0.54 mag in V, 0.58 mag in I, 0.51 mag in J and 0.21 mag in H for the galaxies in our sample.

\begin{figure*}
\centering
    \includegraphics[width=\textwidth]{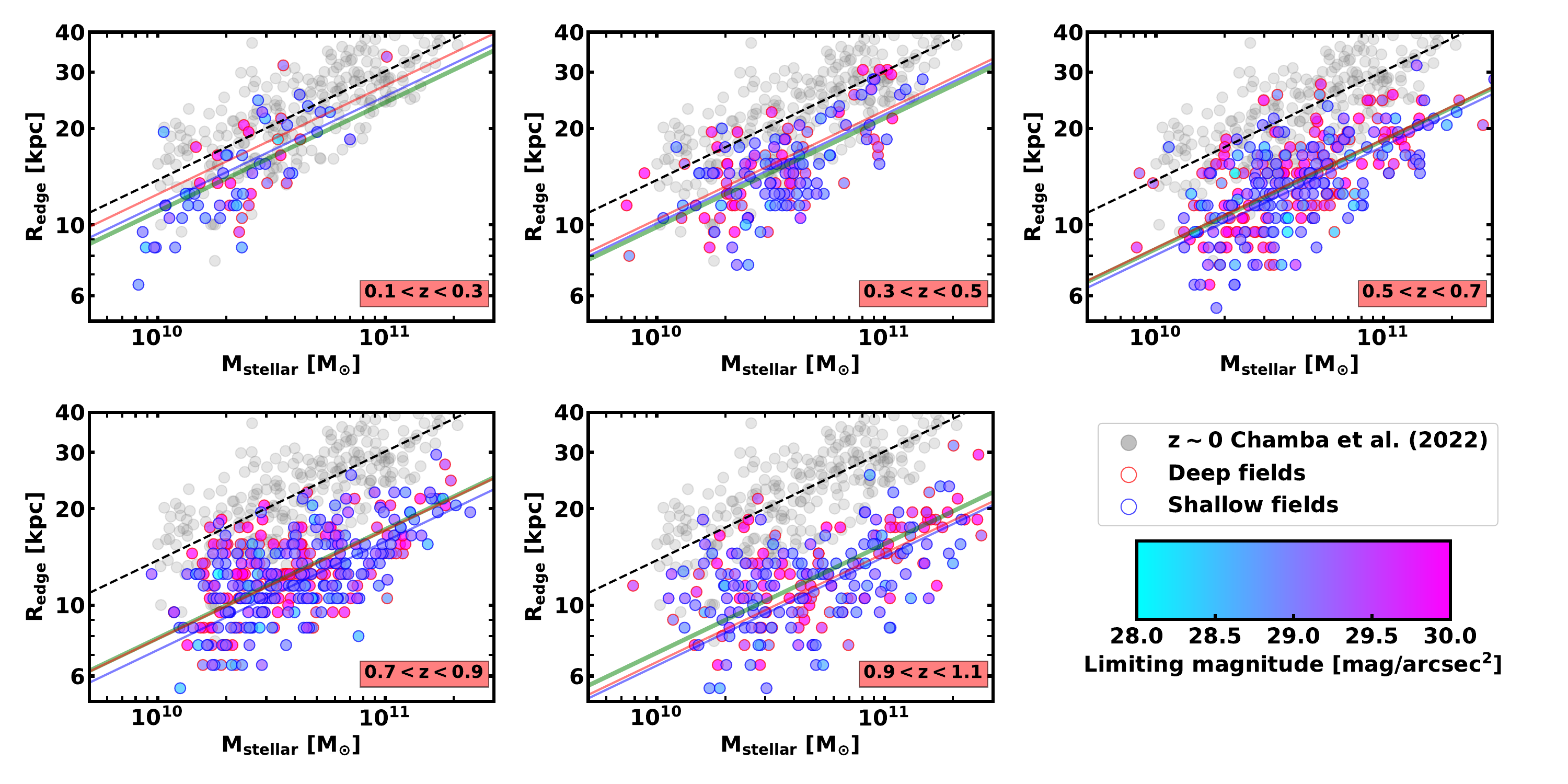}
    \caption{Size - mass  relation of our sample splitting the galaxy sample according to the image depth on the I-band (F814W filter). We separated those objects that belong to the deeper fields (GOODS South and North) with those from the shallower fields (COSMOS, UDS and EGS). We present the global fits to the entire sample (in green, with the black dashed line corresponding to the local sample) together with those to the deep and shallow galaxy subsamples (red and blue lines respectively). The depth of the different data does not affect to the size evolution of the objects.}
    \label{fig:trunc_mass_rel_depth}
\end{figure*}

The result of splitting our sample in two different groups depending on their observed surface brightness limiting depth is shown in Fig. \ref{fig:trunc_mass_rel_depth}. Although the images we have used can differ up two magnitudes in limiting surface brightness, this has not visible consequences on the size-mass relationship we retrieve. To further explore this issue, as stated in Section \ref{subsec:mass-sigma}, a Kolmogorov-Smirnov test using the galaxies in the common stellar mass range 3 to 7$\times$10$^{10}$ M$_{\odot}$ (at all redshift bins) concluded that both subsamples were indistinguishable.

\section{The effective radius - size relation for the galaxies in our sample}
\label{app:effective_radii}

\begin{figure*}
\centering
    \includegraphics[width=\textwidth]{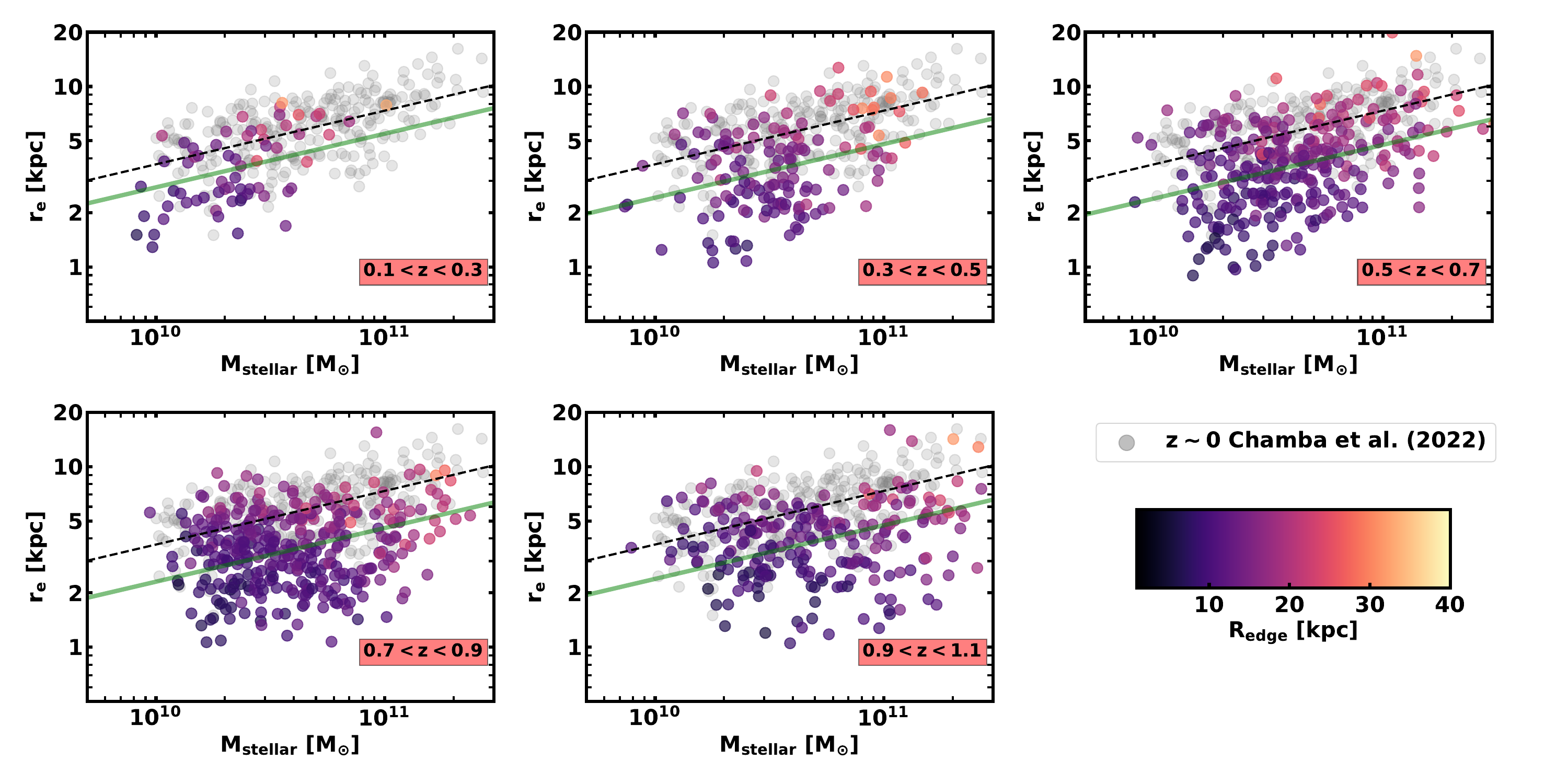}
    \caption{Effective radius - mass relation for the galaxies in our sample. The local sample corresponds to the galaxies used in \citet{Chamba22} while the effective radius of the galaxies explored in this paper were retrieved from  \citet{vanderWel12}. The galaxies are colour coded according to the size measured in this work  (R$_{\rm edge}$). The dashed lines correspond to the fitting to the local sample while the green solid lines to the fits to the higher redshift galaxies. The slope of the global fitting is left fixed to the local sample. There is a mild decrease of the effective radii with redshift, but not as steep as in the R$_{\rm edge}$'s case. Besides, the scatter ($\sim$0.2 dex) is $\sim$2 times larger than the one found using  R$_{\rm edge}$ as a size indicator ($\sim$0.1 dex).}
    \label{fig:mass_size_rel_effective_radii}
\end{figure*}

It is worth exploring what the effective radius - mass relation is for the galaxies in our sample given the fact that the effective radius parameter is the most common size proxy for measuring the galaxy size in the last two decades. We show it in Fig. \ref{fig:mass_size_rel_effective_radii}. As found in previous studies \citep[see e.g.][]{vanderWel14}, the  evolution of the effective radius at a fixed stellar mass for disc galaxies is very mild since z=1. Moreover, the scatter is significantly larger ($\sim$0.2 dex) compared to the observed size-mass relation using the extension of the star forming disc (R$_{\rm edge}$) as a proxy for measuring the global size of the galaxies.

\section{The evolution of the central surface stellar mass densities with redshift}
\label{app:central_mass_profiles}

\begin{figure*}
\centering
    \includegraphics[width=\textwidth]{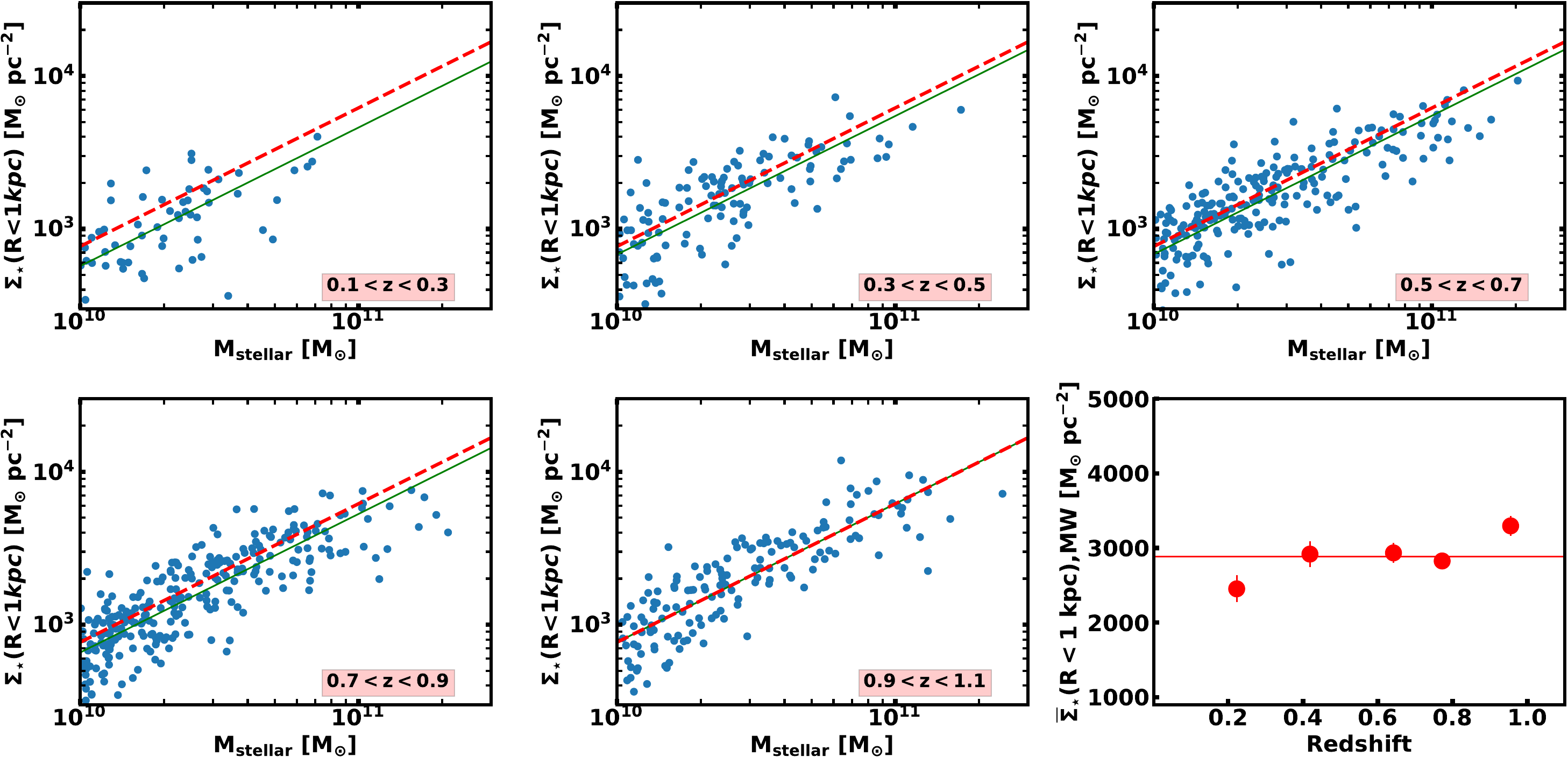}
    \caption{Relation between the central  stellar mass surface density of our disc galaxies (measured by the mean surface mass density within 1 kpc) and the stellar mass. The red dashed line corresponds to the fit to the data points in the highest redshift bin (0.9 $<$ z $<$ 1.1). The green lines at each redshift bin are the fits to the data points using a fixed slope equal to that measured at the highest redshift bin. The lower right panel shows the evolution with redshift of the central surface mass density of galaxies of similar mass to the Milky Way (i.e. M$_{\rm stellar}$ = 5$\times$10$^{10}$ M$_{\odot}$). Errors come from bootstrapping 10$^4$ times our measurements. In contrast to the evolution of the stellar surface mass density at the edge of the galaxy, the central density remains relatively unchanged over the last 8 Gyr.}
    \label{fig:central_mass_densities}
\end{figure*}

While the evolution of the stellar  mass surface density at the edge changes by about an order of magnitude since z=1, we have quantified the change in the central part of disc galaxies over the same redshift interval. Since this part of the galaxy is  thought to have formed first (unlike the outer part of the galaxy where the edge is located), we expect it to change very little (if at all) with redshift. To make a quantitative analysis of this quantity, we have measured the average surface mass density within 1 kpc. The evolution of the central surface stellar mass density can be seen in Fig. \ref{fig:central_mass_densities}. The red dashed line corresponds to the fit to the data points in the highest redshift bin. The green dashed lines at each redshift bin are the fits to the data points using a fixed slope equal to that measured at the highest redshift bin. The lower right panel shows the evolution with redshift of the central surface mass density of galaxies of similar mass to the Milky Way (i.e. M$_{\rm stellar}$ = 5$\times$10$^{10}$ M$_{\odot}$). Errors come from bootstrapping 10$^4$ times our measurements. As expected, the central surface mass density is fairly constant with redshift (with an evolution of <1.3 since z=1).

\end{appendix}

\end{document}